%% file: main.tex
\documentclass[sigplan,screen,nonacm,10pt]{acmart}

\def\BibTeX{{\rm B\kern-.05em{\sc i\kern-.025em b}\kern-.08emT\kern-.1667em\lower.7ex\hbox{E}\kern-.125emX}}

\renewcommand\footnotetextcopyrightpermission[1]{}

\newcommand{\resultswebsiteurl}{\url{scalable20.mlmodelscope.org}\xspace}

\setcopyright{none}

\usepackage{hyphenat}
\usepackage{balance}

\usepackage{microtype}
\usepackage{graphicx}
\usepackage{verbatim}
\usepackage{booktabs} %
\usepackage{soul}
\usepackage{amsthm}
\usepackage{amsmath}
\usepackage{blindtext}
\usepackage{fancyhdr}
\usepackage{arydshln}
\usepackage{multirow}
\usepackage{xspace}
\usepackage{array}
\usepackage{siunitx}
\usepackage{makecell}
\usepackage{rotating}
\usepackage{filecontents}
\usepackage{pgfplots}
\usepackage{pgfplotstable}
\usepackage{tikz}
\usepackage[normalem]{ulem}
\usepackage{enumitem}
\usepackage{listings}
\usepackage{xargs}                      %
\usepackage{amssymb}%
\usepackage{pifont}%
\usepackage{tcolorbox}
\usepackage{environ}

\usepackage{caption}
\usepackage{subcaption}

\makeatletter
\newsavebox{\measure@tikzpicture}
\NewEnviron{scaletikzpicturetowidth}[1]{%
  \def\tikz@width{#1}%
  \begin{lrbox}{\measure@tikzpicture}%
  \BODY
  \end{lrbox}%
  \pgfmathparse{#1/\wd\measure@tikzpicture}%
  \BODY
}

\setlist{noitemsep,nolistsep}

\newcommand{\cmmnt}[1]{\ignorespaces}

\definecolor{myred}{rgb}{0.843137,0.188235,0.152941}
\definecolor{myblack}{rgb}{0.27451,0.32549,0.384314}
\definecolor{mygreen}{rgb}{0.301961,0.686275,0.290196}
\definecolor{myyellow}{rgb}{0.996078,0.878431,0.564706}
\definecolor{myblue}{rgb}{0.568627,0.74902,0.858824}

\pgfplotsset{compat=newest,}
\usetikzlibrary{babel}

\pgfplotsset{every axis/.style={scale only axis}}

\input{latex_tools/colors}

\usepgfplotslibrary{colorbrewer}
\pgfplotsset{cycle list/Dark2-8}

\newtcolorbox{observationbox}[1][]
{
    breakable,
    left=1pt,
    right=1pt,
    top=1pt,
    bottom=1pt,
    colback=gray!20,
    colframe=black,
    width=\dimexpr\columnwidth\relax,
    enlarge left by=0mm,
    boxsep=5pt,
    arc=0pt,outer arc=0pt,
    #1
}

\definecolor{lightyellow}{RGB}{255, 250, 236}
\definecolor{textdark}{RGB}{100, 52, 20}
\definecolor{borderorange}{RGB}{253, 129, 36}
\definecolor{lightgray}{RGB}{214, 214, 214}
\definecolor{countrygray}{RGB}{153, 153, 153}
\definecolor{circlered}{RGB}{176, 0, 29}
\definecolor{circlegreen}{RGB}{36, 94, 1}

\newcommand{\carml}{MLModelScope\xspace}

\newcommand{\ignore}[1]{}

\DeclareRobustCommand*\feature[1]{\tikz[baseline=(char.base)]{
            \node[rounded corners,white,draw,fill=black,line width=1pt,rounded corners=2pt,inner sep=1.4pt] (char) {\normalfont\sffamily\textsf{F#1}};}}
\DeclareRobustCommand*\kernel[1]{\tikz[baseline=(char.base)]{
            \node[rounded corners,white,draw,fill=black,line width=1pt,rounded corners=2pt,inner sep=1.4pt] (char) {\normalfont\sffamily\textsf{K#1}};}}

\newtcbox\questionbox{hbox, on line, colback=black, enhanced, frame hidden, boxrule=0pt,
    top=-2pt, bottom=-2pt, right=-2pt, left=-2pt, rounded corners, arc=2pt}

\DeclareRobustCommand*\circledwhitegreen[1]{\tikz[baseline=(char.base)]{
            \node[shape=circle,line width=0.5mm,circlegreen,draw,text=black,inner sep=0.5pt,anchor=base] (char) {\normalfont\sffamily\bfseries\footnotesize{#1}};}}
\DeclareRobustCommand*\circledwhitered[1]{\tikz[baseline=(char.base)]{
            \node[shape=circle,line width=0.5mm,circlered,draw,text=black,inner sep=0.5pt,anchor=base] (char) {\normalfont\sffamily\bfseries\footnotesize{#1}};}}
\DeclareRobustCommand*\circledwhite[1]{\tikz[baseline=(char.base)]{
            \node[shape=circle,line width=0.5mm,draw,inner sep=0.5pt,anchor=base] (char) {\normalfont\sffamily\bfseries\footnotesize{#1}};}}

\newcommand*\featureReproducible{\feature{1}}
\newcommand*\featureConsistent{\feature{2}}
\newcommand*\featureAgnostic{\feature{3}}
\newcommand*\featureScalable{\feature{4}}
\newcommand*\featureVersion{\feature{5}}
\newcommand*\featureWorkflow{\feature{6}}
\newcommand*\featureBenchScenarios{\feature{7}}
\newcommand*\featureReport{\feature{8}}
\newcommand*\featureInspection{\feature{9}}
\newcommand*\featureUI{\feature{10}}

\newcommand{\FORGET}[1]{}

\input{latex_tools/yaml}
\input{latex_tools/golang}
\input{latex_tools/protobuf}  %

\hypersetup{
    colorlinks=false,
    linkcolor=black
}

\title{The Design and Implementation of a Scalable DL Benchmarking Platform}

\newcolumntype{C}{>{\centering\arraybackslash} m{.1\linewidth} }  %

\makeatletter

\newcommand\addauthornote[1]{%
  \if@ACM@anonymous\else
    \g@addto@macro\addresses{\@addauthornotemark{#1}}%
  \fi}

\newcommand\@addauthornotemark[1]{\let\@tmpcnta\c@footnote
   \setcounter{footnote}{#1}\addtocounter{footnote}{-1}
    \g@addto@macro\@currentauthors{\footnotemark\relax\let\c@footnote\@tmpcnta}}

\makeatother

\begin{document}

\author{Cheng Li}
\addauthornote{1}
\affiliation{%
  \institution{University of Illinois Urbana-Champaign}
  \city{Urbana}
  \state{Illinois}
}
\email{cli99@illinois.edu}

\author{Abdul Dakkak}
\addauthornote{1}
\affiliation{%
  \institution{University of Illinois Urbana-Champaign}
  \city{Urbana}
  \state{Illinois}
}
\email{dakkak@illinois.edu}

\author{Jinjun Xiong}
\affiliation{%
  \institution{IBM T. J. Watson Research Center}
  \city{Yorktown Heights}
  \state{New York}
}
\email{jinjun@us.ibm.com}

\author{Wen-mei Hwu}
\affiliation{%
  \institution{University of Illinois Urbana-Champaign}
  \city{Urbana}
  \state{Illinois}
}
\email{w-hwu@illinois.edu}

\renewcommand{\shortauthors}{Cheng Li, Abdul Dakkak, et al.}

\begin{abstract}

\input{sec/0-abstract.tex}

\end{abstract}

\maketitle

\footnotetext[1]{The two authors contributed equally to this paper.}

\input{latex_figures/fig_data.tex}
\input{sec/1-intro.tex}

\input{sec/2-background.tex}

\input{sec/3-objectives.tex}

\input{sec/4-design.tex}

\input{sec/5-implementation.tex}

\input{sec/6-evaluation.tex}

\input{sec/2.x-related.tex}

\input{sec/7-conclusion.tex}

\input{sec/99-ack.tex}

\bibliographystyle{ACM-Reference-Format}
\bibliography{main}

\end{document}

%% file: latex_tools/colors.tex
\definecolor{plotcolor1}{rgb}{0.568627,0.74902,0.858824}
\definecolor{plotcolor2}{rgb}{0.996078,0.878431,0.564706}
\definecolor{plotcolor3}{rgb}{0.27451,0.32549,0.384314}
\definecolor{plotcolor4}{rgb}{0.843137,0.188235,0.152941}
\definecolor{plotcolor5}{rgb}{0.988235,0.552941,0.34902}
\definecolor{plotcolor6}{rgb}{0.596078,0.305882,0.639216}
\definecolor{plotcolor7}{rgb}{0.65098,0.337255,0.156863}
\definecolor{plotcolor8}{rgb}{0.105882,0.619608,0.466667}
\definecolor{plotcolor9}{rgb}{1.,1.,0.6}
\definecolor{plotcolor10}{rgb}{0.745098,0.729412,0.854902}
\definecolor{plotcolor11}{rgb}{0.984314,0.501961,0.447059}
\definecolor{plotcolor12}{rgb}{0.501961,0.694118,0.827451}
\definecolor{plotcolor13}{rgb}{1.,1.,0.2}
\definecolor{plotcolor14}{rgb}{0.992157,0.705882,0.384314}
\definecolor{plotcolor15}{rgb}{0.988235,0.803922,0.898039}
\definecolor{plotcolor16}{rgb}{0.701961,0.870588,0.411765}
\definecolor{plotcolor17}{rgb}{0.215686,0.494118,0.721569}
\definecolor{plotcolor18}{rgb}{0.941176,0.231373,0.12549}
\definecolor{plotcolor19}{rgb}{0.168627,0.54902,0.745098}
\definecolor{plotcolor20}{rgb}{0.552941,0.827451,0.780392}
\definecolor{plotcolor21}{rgb}{0.968627,0.505882,0.74902}
\definecolor{plotcolor22}{rgb}{0.6,0.6,0.6}
\definecolor{plotcolor23}{rgb}{0.301961,0.686275,0.290196}
\definecolor{plotcolor24}{rgb}{0.980392,0.501961,0.447059}

%% file: latex_tools/yaml.tex
\definecolor{yamlred}{rgb}{0.843137,0.188235,0.152941}
\definecolor{yamlblue}{rgb}{0.27451,0.32549,0.384314}
\definecolor{yamlkey}{rgb}{0.980392,0.501961,0.447059}

\newcommand\YAMLfontsize{\fontsize{6}{6}}
\newcommand\YAMLcolonstyle{\color{black}\mdseries\YAMLfontsize\ttfamily}
\newcommand\YAMLkeystyle{\bfseries\color{yamlblue}\YAMLfontsize\ttfamily}
\newcommand\YAMLvaluestyle{\color{black}\mdseries\YAMLfontsize\ttfamily}

\definecolor{yamlgray}{rgb}{0.5,0.5,0.5}
\lstdefinelanguage{yaml}
{
  keywords={true,false,null,y,n},
  keywordstyle=\YAMLvaluestyle,
  basicstyle=\YAMLfontsize\YAMLkeystyle,                                 %
  sensitive=false,
  comment=[l]{\#},
  morecomment=[s]{/*}{*/},
  commentstyle=\YAMLcolonstyle\color{yamlred}\ttfamily\YAMLfontsize\ttfamily,
  stringstyle=\YAMLvaluestyle\ttfamily\YAMLfontsize\ttfamily,
  moredelim=[l][\color{orange}]{\&},
  moredelim=[l][\color{yamlred}]{*},
  moredelim=**[il][\YAMLcolonstyle{:}\YAMLvaluestyle]{:},   %
  morestring=[b]',
  morestring=[b]",
  literate =    {---}{{\ProcessThreeDashes}}3   
                {|}{{\textcolor{yamlred}\textbar}}1 
                {\ -\ }{{\mdseries\ -\ }}3,       
    frame=top,
    frame=bottom,
    numbers=left,
    rulesep=1pt,
	breaklines=true,
    mathescape=true,
    xleftmargin=2em,
    framexleftmargin=3em,
    stepnumber=1,
    escapechar=|,
    showstringspaces=false,
    numberstyle=\tiny\color{yamlgray},
}

\makeatother

\newcommand\ProcessThreeDashes{\llap{\color{cyan}\mdseries-{-}-}}

%% file: latex_tools/golang.tex
\usepackage{xcolor}
\usepackage{textcomp}

\definecolor{comment}{RGB}{0,128,0}    %
\definecolor{string}{RGB}{255,0,0}     %
\definecolor{keyword}{RGB}{0,0,255}    %
\definecolor{function}{RGB}{128,0,128} %
\definecolor{type}{RGB}{128,0,0}       %

\lstdefinestyle{go}{
	commentstyle=\color{comment},
	stringstyle=\color{string},
	keywordstyle=[1]\color{keyword},
	keywordstyle=[2]\color{function},
	keywordstyle=[3]\color{type},
	basicstyle=\footnotesize\ttfamily,
	numbers=left,
	numberstyle=\tiny,
	numbersep=5pt,
	frame=lines,
	breaklines=true,
	prebreak=\raisebox{0ex}[0ex][0ex]{\ensuremath{\hookleftarrow}},
	showstringspaces=false,
	upquote=true,
	tabsize=3,
}

\lstdefinelanguage{go}{
	sensitive=true,
	alsoletter={\%},
	comment=[l]{//},
	morecomment=[s]{/*}{*/},
	string=[b]{"},
	morestring=[b]{`},
	morestring=[b]{'},
	keywords=[1]{
break, case, chan, const, continue, default, defer, else, fallthrough, for,
func, go, goto, if, import, interface, map, package, range, return, select,
struct, switch, type, var
	},
	morekeywords=[2]{
append, cap, close, complex, copy, delete, imag, len, make, new, panic, print,
println, real, recover
	},
	morekeywords=[3]{
bool, byte, complex64, complex128, error, float32, float64, int, int8, int16,
int32, int64, rune, string, uint, uint8, uint16, uint32, uint64, uintptr
	},
}

%% file: latex_tools/protobuf.tex
\usepackage{xcolor}
\usepackage{xcolor-solarized}
\usepackage{textcomp}

\definecolor{protored}{rgb}{0.843137,0.188235,0.152941}
\definecolor{protoblue}{rgb}{0.27451,0.32549,0.384314}
\definecolor{protokey}{rgb}{0.980392,0.501961,0.447059}
\definecolor{protogray}{rgb}{0.8,0.8,0.8}

\definecolor{proto_basic}{rgb}{0,0,0}          
\definecolor{proto_keyword}{rgb}{0.27451,0.32549,0.384314}  
\definecolor{proto_type}{rgb}{0.843137,0.188235,0.152941} 
\definecolor{proto_options}{RGB}{128,0,128} 
\definecolor{proto_comment}{RGB}{128,128,128} 
\definecolor{proto_string}{rgb}{0.843137,0.188235,0.152941}          %
\definecolor{proto_number}{RGB}{108,113,196}  
\definecolor{proto_ident}{RGB}{0,0,0}         
\definecolor{proto_digits}{rgb}{0.27451,0.32549,0.384314}     
\definecolor{proto_background}{RGB}{255,255,255}  

\newcommand\protoFontsize{\fontsize{6}{6}}
\newcommand\protoColonstyle{\color{black}\mdseries\protoFontsize\ttfamily}

\lstdefinestyle{protobuf}{
  basicstyle=\protoFontsize\color{proto_basic}\protoColonstyle,
	keywordstyle=[1]\protoColonstyle\color{proto_keyword},
	keywordstyle=[2]\protoColonstyle\color{proto_type},
	keywordstyle=[3]\protoColonstyle\color{proto_options},
	commentstyle= \protoColonstyle\color{proto_comment},
	stringstyle=\color{proto_string},
  identifierstyle=\color{proto_ident},
	breaklines=true,
	showstringspaces=false,
	tabsize=2,
  numberstyle=\protoColonstyle\color{protogray},
  literate={0}{{\textcolor{proto_digits}{0}}}{1}%
           {1}{{\textcolor{proto_digits}{1}}}{1}%
           {2}{{\textcolor{proto_digits}{2}}}{1}%
           {3}{{\textcolor{proto_digits}{3}}}{1}%
           {4}{{\textcolor{proto_digits}{4}}}{1}%
           {5}{{\textcolor{proto_digits}{5}}}{1}%
           {6}{{\textcolor{proto_digits}{6}}}{1}%
           {7}{{\textcolor{proto_digits}{7}}}{1}%
           {8}{{\textcolor{proto_digits}{8}}}{1}%
           {9}{{\textcolor{proto_digits}{9}}}{1}%
           {.0}{{\textcolor{proto_digits}{.0}}}{2}%
           {.1}{{\textcolor{proto_digits}{.1}}}{2}%
           {.2}{{\textcolor{proto_digits}{.2}}}{2}%
           {.3}{{\textcolor{proto_digits}{.3}}}{2}%
           {.4}{{\textcolor{proto_digits}{.4}}}{2}%
           {.5}{{\textcolor{proto_digits}{.5}}}{2}%
           {.6}{{\textcolor{proto_digits}{.6}}}{2}%
           {.7}{{\textcolor{proto_digits}{.7}}}{2}%
           {.8}{{\textcolor{proto_digits}{.8}}}{2}%
           {.9}{{\textcolor{proto_digits}{.9}}}{2}%
           {int32}{{\textcolor{proto_type}{int32}}}{5}%
           {int64}{{\textcolor{proto_type}{int64}}}{5}%
           {uint32}{{\textcolor{proto_type}{uint32}}}{6}%
           {uint64}{{\textcolor{proto_type}{uint64}}}{6}%
           {sint32}{{\textcolor{proto_type}{sint32}}}{6}%
           {sint64}{{\textcolor{proto_type}{sint64}}}{6}%
           {fixed32}{{\textcolor{proto_type}{fixed32}}}{7}%
           {fixed64}{{\textcolor{proto_type}{fixed64}}}{7}%
           {sfixed32}{{\textcolor{proto_type}{sfixed32}}}{8}%
           {sfixed64}{{\textcolor{proto_type}{sfixed64}}}{8}%
           {java_string_check_utf8}{{%
             \textcolor{proto_options}{java_string_check_utf8}}}{2}%
           {\ }{{ }}{1},
	prebreak=\raisebox{0ex}[0ex][0ex]{\ensuremath{\hookleftarrow}},
	upquote=true,
}

\definecolor{pbgray}{rgb}{0.5,0.5,0.5}
\lstdefinelanguage[2]{protobuf}{%
    sensitive=true,%
    numbers=left,
	breaklines=true,
    mathescape=true,
    xleftmargin=2em,
    framexleftmargin=3em,
    stepnumber=1,
    escapechar=|,
    numberstyle=\protoColonstyle\color{pbgray},  
    frame=top,
    frame=bottom,
    morecomment=[l]{//},%
    morestring=[b]{"},%
    morekeywords={enum,oneof,map,syntax,public,import,option,package,message,%
        group,optional,required,repeated,default,reserved,extend,extensions,%
        to,max,service,rpc,returns,true,false},%
    morekeywords=[2]{%
        double,float,int32,int64,uint32,uint64,sint32,sint64,%
        fixed32,fixed64,sfixed32,sfixed64,bool,string,bytes},%
    morekeywords=[3]{%
        deprecated, uninterpreted_option,%
        java_package,java_outer_classname,java_multiple_files,%
        java_generate_equals_and_hash,java_string_check_utf8,optimize_for,%
        go_package,cc_generic_services,java_generic_services,%
        py_generic_services,cc_enable_arenas,obj_class_prefix,%
        csharp_namespace,%
        message_set_wire_format,no_standard_descriptor_accessor,map_entry,%
        ctype, packed,jstype,lazy,weak,%
        allow_alias}%
}
\lstalias[]{protobuf2}[2]{protobuf}

\lstdefinelanguage[3]{protobuf}[2]{protobuf}{%
    deletekeywords={
        group,%
        extensions, to, extend, max,%
        required,%
        optional,%
        default}%
}
\lstalias[]{protobuf3}[3]{protobuf}

%% file: sec/0-abstract.tex
The current Deep Learning (DL) landscape is fast-paced 
and is rife with non-uniform models, hardware/software (HW/SW) stacks, but lacks a DL benchmarking platform to facilitate evaluation and comparison of DL innovations, be it models, frameworks, libraries, or hardware. 
Due to the lack of a benchmarking platform, the current practice of evaluating the benefits of proposed DL innovations is both arduous and error-prone --- stifling the adoption of the innovations.

In this work, we first identify $10$ design features which are desirable within a DL benchmarking platform.
These features include: performing the evaluation in a consistent, reproducible, and scalable manner, being framework and hardware agnostic, supporting real-world benchmarking workloads, providing in-depth model execution inspection across the HW/SW stack levels, etc.
We then propose \carml, a DL benchmarking platform design that realizes the $10$ objectives.
\carml proposes a specification to define DL model evaluations and techniques to provision the evaluation workflow using the user-specified HW/SW stack.
\carml defines abstractions for frameworks and supports board range of DL models and evaluation scenarios.
We implement \carml as an open-source project with support for all major frameworks and hardware architectures.
Through \carml's evaluation and automated analysis workflows, we performed case-study analyses of $37$ models across $4$ systems and show how model, hardware, and framework selection affects model accuracy and performance under different benchmarking scenarios.
We further demonstrated how \carml's tracing capability gives a holistic view of model execution and helps pinpoint bottlenecks.

%% file: sec/1-intro.tex
\section{Introduction}\label{sec:intro}

The emergence of Deep Learning (DL) as a popular application domain has led to many innovations.
Every day, diverse DL models, as well as hardware/software (HW/SW) solutions, are proposed --- be it algorithms, frameworks, libraries, compilers, or hardware.
DL innovations are introduced at such a rapid pace~\citep{dean2018new} that being able to evaluate and compare these innovations in a timely manner is critical for their adoption.
As a result, there have been concerted community efforts in developing DL benchmark suites~\cite{mlperf,aimatrix,dawnbench,gao2019aibench} where common models are selected and curated as benchmarks.

DL benchmark suites require significant effort to develop and maintain and thus have limited coverage of models (usually a few models are chosen to represent a DL task).
Within these benchmark suites, model benchmarks are often developed independently as a set of ad-hoc scripts.
To consistently evaluate two models requires one to use the same evaluation code and HW/SW environment.
Since the model benchmarks are ad-hoc scripts, a fair comparison requires a non-trivial amount of effort.
Furthermore, DL benchmarking often requires evaluating models across different combinations of HW/SW stacks.
As HW/SW stacks are increasingly being proposed, there is an urging need for a DL benchmarking platform that consistently evaluates and compares different DL models across HW/SW stacks, while coping with the fast-paced and diverse landscape of DL.

As a fledgling field, the benchmarking platform design for DL faces new challenges and requirements.
DL model evaluation is a complex process where the model and HW/SW stack must work in unison, and the benefit of a DL innovation is dependent on this interplay.
Currently, there is no standard to specify or provision DL evaluations, and reproducibility is a significant ``pain-point'' within the DL community~\citep{plesser2018reproducibility,ghanta2018systems,li2019random}.
Thus, the benchmarking platform design must guarantee a \featureReproducible{}~\textbf{reproducible evaluation} along with \textbf{\featureConsistent{}~consistent evaluation}.

Aside from \feature{1-2}, the design should: 
be \textbf{\featureAgnostic{} frameworks and hardware agnostic} to support model evaluation using diverse HW/SW stacks;
be capable of performing \featureScalable{}~\textbf{scalable evaluation} across systems to cope with the large number of evaluations due to the diverse model/HW/SW combinations;
support different \featureBenchScenarios{}~\textbf{benchmarking scenarios}  which mimic the real-world workload exhibited in online, offline, and interactive applications;
have a \featureReport{}~\textbf{benchmarking analysis and reporting} workflow which analyzes benchmarking results across evaluation runs and generates summary reports; 
enable \featureInspection{}~\textbf{model execution inspection} to identify bottlenecks within a model-, framework-, and system-level components.
Other features such as: \featureVersion{}~\textbf{artifact versioning}, \featureWorkflow{}~\textbf{efficient evaluation workflow},  and \featureUI{}~\textbf{different user interfaces} are also desirable to increase the design's scalability and usability.
We discuss the design objectives in detail in Section~\ref{sec:objectives}.

In this paper, we propose \textit{\carml}, a scalable DL benchmarking platform design that realizes the above $10$ objectives and facilitates benchmarking, comparison, and understanding of DL model executions.
\carml{} achieves the design objectives by
proposing a specification to define DL model evaluations;
introducing techniques to consume the specification and provision the evaluation workflow using the specified HW/SW stack;
using a distributed scheme to manage, schedule, and handle model evaluation requests;
supporting pluggable workload generators;
defining common abstraction API across frameworks;
providing across-stack tracing capability that allows users to inspect model execution at different HW/SW abstraction levels;
defining an automated evaluation analysis workflow for analyzing and reporting evaluation results;
and, finally, exposing the capabilities through a web and command-line interface.

We implement \carml{} and integrate it with the Caffe~\cite{caffe}, Caffe2~\cite{caffe2}, CNTK~\cite{cntk}, MXNet~\cite{mxnet}, PyTorch~\cite{pytorch}, TensorFlow~\cite{tensorflow}, TensorFlow  Lite~\cite{tflite}, and TensorRT~\cite{tensorrt} frameworks.
\carml runs on ARM, PowerPC, and x86 and supports CPU, GPU, and FPGA execution.
We bootstrap \carml with over $300$ built-in models covering different DL tasks such as image classification, object detection, semantic segmentation, etc.
\carml{} is open-source, extensible, and customizable ---  coping with the fast-paced DL landscape.
To the authors' knowledge, this paper is the first to describe the design and implementation of a scalable DL benchmarking platform.

We showcase \carml's benchmarking, inspection, and analysis capabilities using several case studies.
We use \carml to evaluate $37$ DL models and systematically compare their performance using $4$ systems under different benchmarking scenarios.
We perform comparisons to understand the correlation between a model's accuracy, its size, and its achieved latency and maximum throughput.
We then use \carml's tracing capability to identify the bottlenecks of the evaluation and use its ``zoom-in'' feature to inspect the model execution at different HW/SW levels.
We demonstrate how, using the analysis workflow, users can easily digest the evaluation results produced by \carml to understand model-, framework-, and system-level bottlenecks.

This paper describes the design and implementation of \carml and is structured as follows.
Section~\ref{sec:background} gives a background.
Section~\ref{sec:objectives} describes the objectives   of \carml.
Section~\ref{sec:design} proposes the \carml design which addresses these objectives and describes its implementation.
Section~\ref{sec:evaluation} performs in-depth evaluations using \carml.
Section~\ref{sec:related} details related work before we conclude in Section~\ref{sec:conclusion}.

%% file: sec/2-background.tex
\section{Background}\label{sec:background}

This section gives a brief background of DL model evaluation and current DL benchmarking practice.

\subsection{DL Model Evaluation Pipeline}

A DL model evaluation pipeline performs input pre-processing, followed by model prediction and output post-processing (Figure~\ref{fig:profile_levels}).
Pre-processing is the process of transforming the user input into a form that can be consumed by the model and post-processing is the process of transforming the model's output to compute metrics.
If we take image classification as an example, the pre-processing step decodes the input image into a tensor of dimensions [$batch$, $height$, $width$, $channel$] ($[N,H,W,C]$), then performs resizing, normalization, etc.
The image classification model's output is a tensor of dimensions $[batch*numClasses]$ which is sorted to get the top $K$ predictions (label with probability).
A DL \textit{model} is defined by its graph topology and its weights.
The graph topology is defined as a set of nodes where each node is a function operator with the implementation provided by a \textit{framework} (e.g. TensorFlow, MXNet, PyTorch).
The framework acts as a ``runtime'' for the model prediction and maps the function operators into \textit{system library} calls.
As can be observed, this pipeline is intricate and has many \textit{levels} of abstraction.
When a slowdown is observed, any one of these levels of abstraction can be suspect.

\subsection{Current DL Benchmarking}

While there has been a drive to provide reference DL benchmarks~\cite{mlperf,gao2019aibench,aimatrix}, the current benchmarking effort is still scattered, lacks a standard benchmarking methodology, and revolves around a series of scripts that evaluate a model on a local system.
To consistently evaluate two models involves: instantiating the same hardware; installing the same software packages and their dependencies;
and, finally, measuring and analyzing the results of both models in the same way.
Because of the use of ad-hoc scripts and lack of a standard way to evaluate models, the above process requires a lot of manual work, and can be error-prone --- often resulting in non-reproducible~\cite{gundersen2018reproducible,ghanta2018systems,li2019random} benchmarking results.
Due to the daunting effort to perform fair benchmarking, 
DL innovations proposed have outpaced researchers' ability to compare and analyze them~\cite{dean2018new}.

%% file: sec/3-objectives.tex
\begin{figure*}
  \centering
  \includegraphics[clip, width=0.90\textwidth]{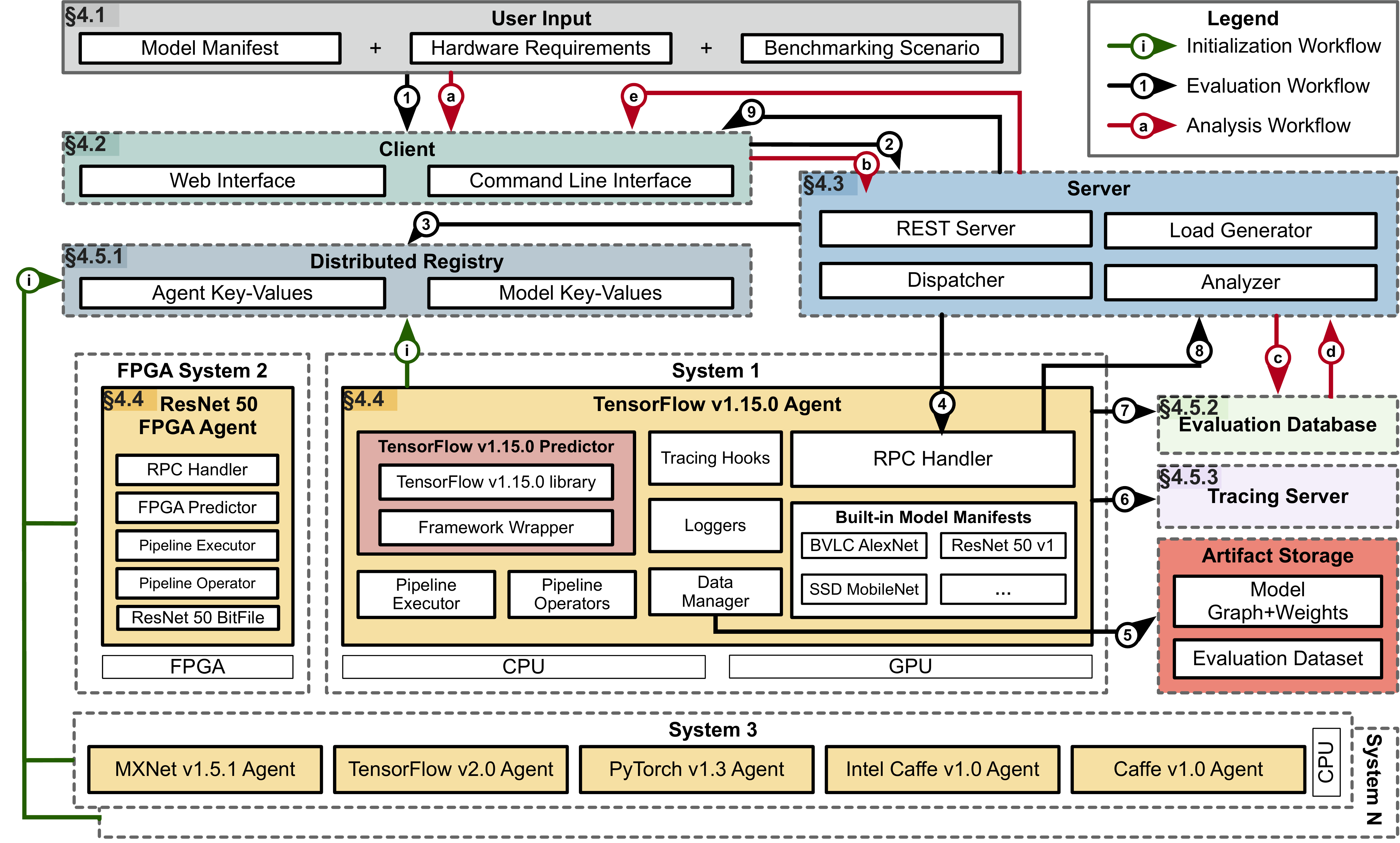}
  \caption{The \carml design and workflows. }
  \label{fig:arch_flow}
\end{figure*}

\section{Design Objectives}\label{sec:objectives}

In this section, we detail $10$ objectives for a DL benchmarking platform design to cope with the fast-evolving DL landscape.
These objectives informed \carml's design choices.

\textbf{\featureReproducible{}~Reproducible Evaluation} ---
Model evaluation is a complex process where the model, dataset, evaluation method, and HW/SW stack must work in unison to maintain the accuracy and performance claims.
Currently, model authors distribute their models and code (usually ad-hoc scripts) by publishing them to public repositories such as GitHub.
Due to the lack of standard specification, model authors may under-specify or omit key aspects of model evaluation.
As a consequence, reproducibility is a ``pain-point'' within the DL community~\cite{gundersen2018reproducible,plesser2018reproducibility,ghanta2018systems, hutson2018artificial,li2019random,tatman2018practical}.
Thus, all aspects of a model evaluation must be specified and provisioned by the platform design to guarantee reproducible evaluation.

\textbf{\featureConsistent{}~Consistent Evaluation} ---
The current practice of publishing models and code also poses challenges to consistent evaluation.
The ad-hoc scripts usually have a tight coupling between model execution and the underlying HW/SW components, making it difficult to quantify or isolate the benefits of an individual component (be it model, framework, or other SW/HW components).
A fair apple-to-apple comparison between model executions requires a consistent evaluation methodology rather than running ad-hoc scripts for each.
Thus the design should have a well-defined benchmarking specification for all models and maximize the common code base that drives model evaluations.

\textbf{\featureAgnostic{}~Framework/Hardware Agnostic} ---
The DL landscape is diverse and there are many DL frameworks (e.g. TensorFlow, MXNet, PyTorch) and hardware (e.g. CPU, GPU, FPGA).
Each has its own use scenarios, features, and performance characteristics.
To have broad support of model evaluation, the design must  support different frameworks and hardware.
Furthermore, the design must be valid without modifications to the frameworks.

\textbf{\featureScalable{}~Scalable Evaluation} ---
DL innovations, such as models, frameworks, libraries, compilers, and hardware accelerators are introduced at a rapid pace~\cite{dean2018new,hazelwood2018applied}.
Being able to quickly evaluate and compare the benefits of DL innovations is critical for their adoption.
Thus the ability to perform DL evaluations with different model/HW/SW setups in parallel and have a centralized management of the benchmarking results is highly desired.
For example, choosing the best hardware out of $N$ candidates for a model is ideally performed in parallel and the results should be automatically gathered for comparison.

\textbf{\featureVersion{}~Artifact Versioning} ---
DL frameworks are continuously updated by the DL community, e.g. the recent versions TensorFlow at the time of writing are v$1.15$ and v$2.0$.
There are many unofficial variants of models, frameworks, and datasets as researchers might update or modify them to suite their respective needs.
To enable management and comparison of model evaluations using different DL artifacts (models, frameworks, and datasets), the artifacts used for evaluation within a benchmarking platform should be versioned.

\textbf{\featureWorkflow{}~Efficient Evaluation Workflow} ---
Before model inference can be performed, the input data has to be loaded into memory and the pre-processing stage transforms it into a form that the model expects.
After the model prediction, the post-processing stage transforms the model's output(s) to a form that can be used to compute metrics.
The input data loading and pre-/post-processing can take a non-negligible amount of time, and become a limiting factor for quick evaluations~\cite{dakkak2019trims}.
Thus the design should handle and process data efficiently in the evaluation workflow.

\textbf{\featureBenchScenarios{}~Benchmarking Scenarios} ---
DL benchmarking is performed under specific scenarios.
These scenarios mimic the usage of DL in online, offline, or interactive applications on mobile, edge, or cloud systems.
The design should support common inference scenarios and be flexible to support custom or emerging workloads as well.

\textbf{\featureReport{}~Benchmarking Analysis and Reporting} ---
Benchmarking produces raw data which needs to be correlated and analyzed to produce human-readable results.
An automated mechanism to summarize and visualize these results within a benchmarking platform can help users quickly understand and compare the results.
Therefore, the design should have a benchmarking result analysis and reporting workflow.

\textbf{\featureInspection{}~Model Execution Inspection} ---
Benchmarking is often followed by performance optimization. However, the complexity of DL model evaluation makes performance debugging challenging as each level within the HW/SW abstraction hierarchy can be a suspect when things go awry.
Current model execution inspection methods rely on the use of a concoction of profiling tools (e.g. Nvidia's Nsight System or Intel's Vtune).
Each profiling tool captures a specific aspect of the HW/SW stack and researchers manually correlate the results to get an across-stack view of the model execution profile.
To ease inspecting model execution bottlenecks, the benchmarking platform design should provide tracing  capability at all levels of HW/SW stack.

\textbf{\featureUI{}~Different User Interfaces} ---
While the command-line is the most common interface in the current benchmarking suites, having other UIs, such as web UI, to accommodate other use cases can greatly boost productivity.
While a command-line interface is often used in scripts to quickly perform combinational evaluations across models, frameworks, and systems, a web UI, on the other hand, can serve as a ``push-button'' solution to benchmarking and provides an intuitive flow for specifying, managing evaluations, and visualizing benchmarking results.
Thus the design should provide UIs for different use cases. %

%% file: sec/4-design.tex
\begin{lstlisting}[
    float=tp,
    floatplacement=tbp,
    language=yaml,
    label={lst:model_manifest},
    escapeinside={(*}{*)},
    captionpos=b,
    caption={The \texttt{MLPerf\_ResNet50\_v1.5}'s model manifest contains all information needed to run the model evaluation using TensorFlow on CPUs or GPUs.},
    escapechar=|
]
|\label{line:start_model_id}|name: MLPerf_ResNet50_v1.5 # model name
|\label{line:end_model_id}|version: 1.0.0 # semantic version of the model 
description: ...
|\label{line:start_framework_id}|framework: # framework information
  name: TensorFlow   
|\label{line:end_framework_id}|  version: '>=1.12.0 <2.0' # framework ver constraint
|\label{line:start_inputs}|inputs: # model inputs
  - type: image  # first input modality
    layer_name: 'input_tensor'
    element_type: float32
    steps: # pre-processing steps
      - decode:
          data_layout: NHWC
          color_mode: RGB
      - resize:
          dimensions: [3, 224, 224]
          method: bilinear
          keep_aspect_ratio: true
      - normalize:
          mean: [123.68, 116.78, 103.94]
|\label{line:end_inputs}|          rescale: 1.0
|\label{line:start_outputs}|outputs: # model outputs
  - type: probability # first output modality
    layer_name: prob
    element_type: float32
    steps: # post-processing steps
      - argsort:
|\label{line:end_outputs}|          labels_url: https://.../synset.txt 
|\label{line:start_proc}|preprocess: [[code]]
|\label{line:end_proc}|postprocess: [[code]]
|\label{line:start_weights}|model: # model sources 
  base_url: https://zenodo.org/record/2535873/files/
  graph_path: resnet50_v1.pb
|\label{line:end_weights}|  checksum: 7b94a2da05d...23a46bc08886
|\label{line:start_attrs}|attributes: # extra model attributes 
  training_dataset:  # dataset used for training
    - name: ImageNet
|\label{line:end_attrs}|    - version: 1.0.0 
\end{lstlisting}

\section{\carml Design and Implementation}\label{sec:design}

We propose \carml, a DL benchmarking platform design that achieves the objectives \feature{1-10} set out in Section~\ref{sec:objectives}.
To achieve \featureScalable~ scalable evaluation, we design \carml as a distributed platform.
To enable \featureBenchScenarios~ real-world benchmarking scenarios, \carml deploys models 
 to be either evaluated using a cloud (as in model serving platforms) or edge (as in local model inference) scenario.
To adapt to the fast pace of DL, \carml is built as a set of extensible and customizable modular components.
We briefly describe each component here and will delve into how they are used later in this section.
Figure~\ref{fig:arch_flow} shows the high level components which include:
\begin{itemize}[nosep,leftmargin=0em,labelwidth=*,align=left]
    \item \textbf{\textit{User Inputs}} --- are the required inputs for model evaluation including: a model manifest (a specification describing how to evaluate a model), a framework manifest (a specification describing the software stack to use), the system requirements (e.g. an X86 system with at least $32$GB of RAM and an NVIDIA V100 GPU), and the benchmarking scenario to employ.
    \item \textbf{\textit{Client}} --- is either the web UI or command-line interface which users use to supply their inputs and initiate the model evaluation by sending a REST request to the \carml server.
    \item \textbf{\textit{Server}} --- acts on the client requests and performs REST API handling, dispatching the model evaluation tasks to \carml agents, generating benchmark workloads based on benchmarking scenarios, and analyzing the evaluation results.
    \item \textbf{\textit{Agents}} ---  runs on different systems of interest and perform model evaluation based on  requests sent by the \carml server.
    Each agent includes logic for downloading model assets, performing input pre-processing, using the framework predictor for inference, and performing post-processing. 
    An agent can be run within a container or as a local process.
    Aside from the framework predictor, all code within an agent is common across frameworks.
    \item \textbf{\textit{Framework Predictor}} --- is a wrapper around a framework and provides a consistent interface across different DL frameworks. The wrapper is designed as a thin abstraction layer so that all DL frameworks can be easily integrated into \carml by exposing a limited number of common APIs.
    \item \textbf{\textit{Middleware}} --- are a set of support services for \carml including: a distributed registry (a key-value store containing entries of  running agents and available models), an evaluation database (a database containing evaluation results), a tracing server (a server to publish profile events captured during an evaluation), and an artifact storage server (a data store repository containing model assets and datasets).
\end{itemize}{}

Figure~\ref{fig:arch_flow} also shows \carml's  three main workflows: \circledwhitegreen{i} initialization, \circledwhite{1-9} evaluation, and \circledwhitered{a-e} analysis.
The initialization workflow is one where all agents self-register by populating the registry with their software stack, system information, and available models for evaluation.
The evaluation workflow works
as follows:
\circledwhite{1} a user inputs the desired model, software and hardware requirements, and benchmarking scenario through a client interface.
The \circledwhite{2} server then accepts the user request, resolves which agents are capable of handling the request by \circledwhite{3} querying the distributed registry, and then \circledwhite{4} dispatches the request to one or more of the resolved agents.
The agent then \circledwhite{5} downloads the required evaluation assets from the artifact storage, performs the evaluation, and \circledwhite{6-7} publishes the evaluation results to the evaluation database and tracing server.
A summary of the results is \circledwhite{8} sent to the server which \circledwhite{9} forwards it to the client.
Finally, the analysis workflow allows a user to perform a more fine-grained and in-depth analysis of results across evaluation runs.
The \carml server handles this workflow by \circledwhitered{a-d}  querying the evaluation database and performing analysis on the results, and \circledwhitered{e} generating a detailed analysis report for the user.
This section describes the \carml components and workflows in detail.

\input{sec/4.0-manifest.tex}

\input{sec/4.4-client.tex}

\input{sec/4.2-server.tex}

\input{sec/4.1-agent.tex}

\input{sec/4.3-middleware.tex}

\input{sec/4.x-extensibility.tex}

%% file: sec/4.0-manifest.tex
\subsection{User Input}\label{sec:spec}

All aspects of DL evaluation --- model, software stack, system, and benchmarking scenario --- must be specified to \carml{} for it to enforce \featureReproducible{} reproducible and \featureConsistent{} consistent evaluation.
To achieve this, \carml{} defines a benchmarking specification covering the $4$ aspects of evaluation.
A model in \carml{} is specified using a \textit{model manifest}, and a software stack is specified using a \textit{framework manifest}.
The manifests are textual specification in YAML~\cite{yaml} format.
The system and benchmarking scenario are user-specified options when the user initiates an evaluation.
The benchmarking specification is not tied to a certain framework or hardware, thus enabling \featureAgnostic{}.
As the model, software stack, system, and benchmarking scenario specification are decoupled, one can easily evaluate the different combinations, enabling \featureScalable{}.
For example, a user can use the same \verb|MLPerf_ResNet50_v1.5| model manifest (shown in Listing~\ref{lst:model_manifest}) to initiate evaluations across different TensorFlow software stacks, systems, and benchmarking scenarios.
To bootstrap the model evaluation process, \carml{} provides built-in model manifests which are embedded in \carml{} agents (Section~\ref{sec:agent}).
For these built-in models, a user can specify the model and framework's name and version in place of the manifest for ease of use.
\carml{} also provides ready-made Docker containers to be used in the framework manifests.
These containers are hosted on Docker hub.

\subsubsection{Model Manifest}\label{sec:manifest}
The model manifest is a text file that specifies information such as the model assets (graph and weights), the pre- and post-processing steps, and other metadata used for evaluation management.
An example model manifest of \texttt{ResNet50 v$1.5$} from MLPerf is shown in Listing~\ref{lst:model_manifest}.
The manifest describes the model name (Lines~\ref{line:start_model_id}-\ref{line:end_model_id}), framework name and version constraint (Lines~\ref{line:start_framework_id}-\ref{line:end_framework_id}), model inputs and pre-processing steps (Lines~\ref{line:start_inputs}-\ref{line:end_inputs}), model outputs and post-processing steps (Lines \ref{line:start_outputs}-\ref{line:end_outputs}), custom pre- and post-processing functions (Lines \ref{line:start_proc}-\ref{line:end_proc}), model assets (Lines \ref{line:start_weights}-\ref{line:end_weights}), and other metadata attributes (Lines \ref{line:start_attrs}-\ref{line:end_attrs}).

\textbf{Framework Constraints} ---
Models are dependent on the framework and possibly the framework version.
Users can specify the framework constraints that a model can execute on.
For example, an ONNX model may work across all frameworks and therefore has no constraint, but other models may only work for TensorFlow versions greater than $1.2.0$ but less than $2$ (e.g. Lines \ref{line:start_framework_id}--\ref{line:end_framework_id} in Listing~\ref{lst:model_manifest}).
This allows \carml{} to support models to target specific versions of a framework and custom frameworks.

\textbf{Pre- and Post-Processing} ---
To perform pre- and post-processing for model evaluation, arbitrary Python functions can be placed within the model manifest (Lines~|\ref{line:start_proc}| and |\ref{line:end_proc}| in Listing~\ref{lst:model_manifest}).
The pre- and post-processing functions are Python functions which have the signature \verb|def fun(env, data)|.
The \verb|env| contains metadata of the user input and \verb|data| is a \verb|PyObject| representation of the user request for pre-process-ing or the model's output for post-processing.
Internally, \carml{} executes the functions within a Python sub-interpret-er~\cite{python-subinterpreter} and passes the data arguments by reference. %
The pre- and post-processing functions are general; i.e. the functions may import external Python modules or download and invoke external scripts.
By allowing arbitrary processing functions, \carml{} works with existing processing codes and is capable of supporting arbitrary input/output modalities.

\textbf{Built-in Pre- and Post-Processing} ---
An alternative way of specifying pre- and post-processing is by defining them as a series of built-in pre- and post-processing pipeline steps (i.e. \textit{pipeline operators}) within the model manifest.
For example, our \carml{} implementation provides common pre-processing image operations (e.g. image decoding, resizing, and normalization) and post-processing operations (e.g. ArgSort, intersection over union, etc.) which are widely used within vision models.
Users can use these built-in operators to define the pre- and post-processing pipelines within the manifest without writing code.
Users define a pipeline by listing the operations within the manifest code (e.g. Lines~\ref{line:start_inputs}--\ref{line:end_inputs} in Listing~\ref{lst:model_manifest} for pre-processing).
The pre- and post-processing steps are executed in the order they are specified in the model manifest. %
The use of built-in processing and function processing pipelines are mutually exclusive.

\textbf{Model Assets} ---
The data required by the model are specified in the model manifest file; i.e. the graph (the \verb|graph_path|) and weights (the \verb|weights_path|) fields.
The model assets can reside within \carml's artifact repository,  on the web, or the local file system of the \carml agent.
If the model assets are remote, then they are downloaded on demand and cached on the local file system.
For frameworks (such as TensorFlow and PyTorch) which use a single file for both the model graph and weights (in deployment), the weights field is omitted from the manifest.
For example, the TensorFlow \texttt{ResNet50 v$1.5$} model assets in Listing~\ref{lst:model_manifest} are stored on the \texttt{Zenodo}~\cite{zenodo} website (Lines~\ref{line:start_weights}-\ref{line:end_weights}) and are downloaded prior to evaluation.

\subsubsection{Framework Manifest \& System Requirements}\label{sec:container}
The framework manifest is a text file that specifies the software stack for model evaluation; an example framework manifest is shown in Listing~\ref{lst:framework_manifest}.
As the core of the software stack, the framework name and version constraints are specified.
To maintain the software stack, and guarantee isolation, the user can further specify docker containers using the \texttt{containers} field.
Multiple containers can be specified to accommodate different systems (e.g. CPU or GPUs).
At the \carml{} initialization phase (\circledwhitegreen{i}),
\carml{} agents (described in Section~\ref{sec:agent}) register themselves by publishing their HW/SW stack information into the distributed registry (described in Section~\ref{sec:registry}).
The \carml server uses this information during the agent resolution process.
The server finds \carml agents satisfying the user's hardware specification and model/framework requirements.
Evaluations are then run on one of (or, at the user request, all of) the agents.
If the user omits the framework manifest in the user input, the \carml server resolves the agent constraints using the model manifest and system information.
This allows \carml to support evaluation on FPGA systems which do not use containers.

\begin{lstlisting}[
    float=tp,
    floatplacement=tbp,
    language=yaml,
    label={lst:framework_manifest},
    escapeinside={(*}{*)},
    captionpos=b,
    caption={An example TensorFlow framework manifest, which contains the software stacks (containers) to run the model evaluation across CPUs or GPUs.},
    escapechar=|
]
name: TensorFlow # framework name
version: 1.15.0 # semantic version of the framework 
description: ...
containers: # containers 
  amd64:
    cpu: carml/tensorflow:1-15-0_amd64-cpu
    gpu: carml/tensorflow:1-15-0_amd64-gpu
  ppc64le:
    cpu: carml/tensorflow:1-15-0_ppc64le-cpu
    gpu: carml/tensorflow:1-15-0_ppc64le-gpu
\end{lstlisting}

\subsubsection{Benchmarking Scenario}

\carml{} provides a set of built-in benchmarking scenarios.
Users pick which scenario to evaluate under.
The benchmarking scenarios include batched inference and online inference with a configurable distribution of time of request (e.g. Poisson distribution of requests). 
The \carml{} server generates an inference request load based on the benchmarking scenario option and sends it to the resolved agent(s) to measure the corresponding benchmarking metrics of the model (detailed in Section~\ref{sec:server}).

%% file: sec/4.4-client.tex
\subsection{\carml Client}\label{sec:ui}

A user initiates a model \circledwhite{1} evacuation or \circledwhitered{a} analysis though the \carml{} \textit{client}.
To enable \featureUI{}, the client can be either a website or a command-line tool that users interact with.
The client communicates with the \carml server through REST API and sends user evaluation requests.
The web user interface allows users to specify a model evaluation through simple clicks and is designed to help users who do not have much DL experience.
For example, for users not familiar with the different models registered, \carml allows users to select models based on the application area --- this lowers the barrier of DL usage. 
The command-line interface is provided for those interested in automating the evaluation and profiling process.
Users can develop other clients that use the REST API to integrate \carml within their AI applications.

%% file: sec/4.2-server.tex
\subsection{\carml Server}\label{sec:server}

The \carml \textit{server} interacts with the \carml client, agent, the middleware.
It uses REST API to communicate with the \carml clients and middleware, and gRPC (Listing~\ref{lst:predict_rpc}) to interact with the \carml{} agents.
To enforce~\featureScalable{}, the \carml server can be load balanced to avoid it being a bottleneck.

In the \circledwhite{1-9} \textit{evaluation workflow}, the server is responsible for \circledwhite{2} accepting tasks from the \carml{} client, \circledwhite{3} querying the distributed registry and resolving the user-specified constraints to find  \carml{} agents capable of evaluating the request, \circledwhite{4} dispatching the evaluation task to the resolved agent(s) and generating loads for the evaluation, \circledwhite{8} collecting the evaluation summary from the agent(s), and \circledwhite{9} returning the result summary to the client.
The load generator is placed on the server to avoid other programs interfering with the evaluation being measured and to emulate real-world scenarios such as cloud serving (\featureBenchScenarios{}). %

In the \circledwhitered{a-e} \textit{analysis workflow}, the server again \circledwhitered{a-b} takes the user input, but, rather than performing evaluation, it \circledwhitered{c} queries the evaluation database (Section ~\ref{sec:database}), and then aggregates and analyzes the evaluation results.
\carml{} enables \featureReport{} through an across-stack analysis pipeline.
It \circledwhitered{d} consumes the benchmarking results and the profiling traces in the evaluation database and performs the analysis.
Then the server \circledwhitered{e} sends the analysis result to the client.
The consistent profiling and automated analysis workflows in \carml allow users to systematically compare across models, frameworks, and system offerings.

%% file: sec/4.1-agent.tex
\begin{lstlisting}[
    float=tp,
    floatplacement=tbp,
    language=protobuf2,
    style=protobuf,
    escapeinside={(*}{*)},
    captionpos=b,
    caption={The predictor interface consists of $3$ API functions.},
    label=lst:predictor_api
]
// Opens a predictor.
ModelHandle ModelLoad(OpenRequest);
// Close an open predictor.
Error ModelUnload(ModelHandle);
// Perform model inference on user data.
PredictResponse Predict(ModelHandle, PredictRequest, PredictOptions);
\end{lstlisting}

\subsection{Agent and Framework Predictor}\label{sec:agent}

A \carml \textit{agent} is a model serving process that is run on a system of interest (within a container or on bare metal) and handles requests from the \carml server.
\carml{} agents continuously listen for jobs and communicate with the \carml{} server through gRPC~\cite{grpc} as shown in Listing~\ref{lst:predict_rpc}.
A \textit{framework predictor} resides within a \carml agent and is a wrapper around a framework and links to the framework's C library.

During the initialization phase (\circledwhitegreen{i}), a \carml{} agent publishes its built-in models and HW/SW information to the \carml distributed registry.
To perform the assigned evaluation task, the agent first \circledwhite{5} downloads the required evaluation assets using the \textit{data manager}, it then executes the model evaluation pipeline which performs the pre-processing, calls the framework's predictor for inference and then preforms the post-processing.
If profiling is enabled, the trace information is published to the \circledwhite{6} tracing server to get aggregated into a single profiling trace. 
 \circledwhite{7 } the benchmarked result and the profiling trace are published to the evaluation database.
Aside from the framework predictor, all the other code --- the data manager, pipeline executor, and tracing hooks --- are shared across agents for different frameworks.
While the default setup of \carml is to run each agent on a separate system, the design does not preclude one from running agents on the same system as separate processes.

\subsubsection{Data Manager}\label{sec:datamanager}

The \textit{data manager} manages the assets (e.g. dataset or model) required by the evaluation as specified within the model manifest.
Assets can be hosted within \carml's \textit{artifact repository}, on the web, or reside in the local file system of the \carml agent.
Both datasets and models are downloaded by the data manager on demand if they are not available on the local system.
If the checksum is specified in the model manifest, the data manager validates the checksum of the asset before using a cached asset or after downloading the asset.
Model assets are stored using the frameworks' corresponding deployment format.
For datasets, \carml supports the use of TensorFlow's TFRecord~\cite{tfrecord} and MXNet's RecordIO~\cite{recordio}.
These dataset formats are optimized for static data and lays out the elements within the dataset as contiguous binary data on disk to achieve better read performance.

\subsubsection{Pipeline Executor and Operators}\label{sec:pipeline}

To enable \featureWorkflow{} efficient evaluation workflow, \carml leverages a streaming data processing pipeline design to perform the model evaluation.
The pipeline is composed of \textit{pipeline operators} which are mapped onto light-weight threads to make efficient use multiple CPUs as well as to overlap I/O with compute.
Each operator within the pipeline forms a producer-consumer relationship by receiving values from the upstream operator(s) (via inbound streams), applies the specified function on the incoming data and usually producing new values, and propagates values downstream (via outbound streams) to the next operator(s).
The pre- and post-processing operations, as well as the model inference, form the operators within the model evaluation pipeline. %

\begin{figure}
    \centering
    \input{latex_figures/fig_language_cpu.tex}
    \input{latex_figures/fig_language_gpu.tex}
    \caption{The \texttt{tf.Session.Run} execution time (normalized to C) across batch sizes for \texttt{Inception-v3} inference on CPU and GPU using TensorFlow with C, Python using NumPy arrays (NumPy), and Python using native lists (Python).}
\label{fig:c_vs_python}
\end{figure}
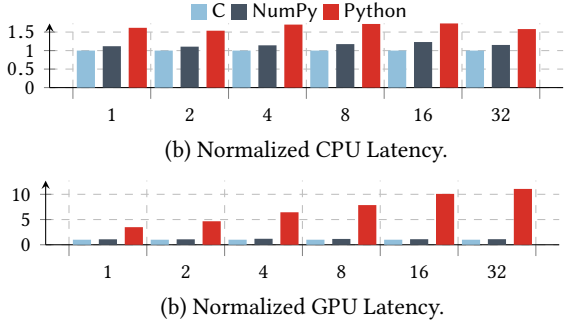

\begin{lstlisting}[
    float=tp,
    floatplacement=tbp,
    language=protobuf2,
    style=protobuf,
    escapeinside={(*}{*)},
    captionpos=b,
    caption={\carml's minimal gRPC interface in protocol buffer format.},
    label=lst:predict_rpc
]
service Predict {
message PredictOptions {
  enum TraceLevel {
    NONE      = 0;
    MODEL     = 1;  // steps in the evaluation pipeline
    FRAMEWORK = 2;  // layers within the framework and above
    SYSTEM    = 3;  // the system profilers and above
    FULL      = 4;  // includes all of the above
  }
  TraceLevel trace_level = 1;
  Options    options     = 2;
}
message OpenRequest {
  string model_name                    = 1;
  string model_version                 = 2;
  string framework_name                = 3;
  string framework_version             = 4;
  string model_manifest                = 5;
  BenchmarkScenario benchmark_scenario = 6;
  PredictOptions    predict_options    = 7;
}
// Opens a predictor and returns a PredictorHandle.
rpc Open(OpenRequest) returns (PredictorHandle){}
// Close a predictor and clear its memory.
rpc Close(PredictorHandle) returns (CloseResponse) {}
// Predict receives a stream of user data and runs
// the predictor on each element of the data according
// to the provided benchmark scenario.
rpc Predict(PredictorHandlePredictorHandle, UserInput) returns (FeaturesResponse) {}
}
\end{lstlisting}

\subsubsection{Framework Predictor}\label{sec:wrapper}

Frameworks provide different APIs (usually across programming languages e.g. C/C++, Python, Java) to perform inference.
To enable \featureConsistent{} consistent evaluation and maximize code reuse, \carml{} wraps each framework's C inference API.
The wrapper is minimal and provides a uniform API across frameworks for performing model loading, unloading, and inference.
This wrapper is called the \textit{predictor interface} and is shown in Listing~\ref{lst:predictor_api}.
\carml{} does not require modifications to a framework and thus pre-compiled binary versions of frameworks (e.g. distributed through Python's pip) or customized versions of a framework work within \carml{}.

\carml is designed to bind to the frameworks' C API to avoid the overhead of using scripting languages.
We demonstrate this overhead by comparing model inference using Python and the C API.
We used TensorFlow \texttt{1.13.0} compiled from source with cuDNN $7.4$ and CUDA Runtime $10.0.1$ on the \texttt{Tesla\_V100} system (Amazon EC2 P3 instance) in Table~\ref{tbl:systems}.
Figure~\ref{fig:c_vs_python} shows the normalized inference latency across language environments on GPUs and CPUs across batch sizes.
On CPU, using Python is $64\%$ and NumPy is $15\%$ slower than C; whereas on GPU Python is $3-11\times$ and NumPy is $10\%$ slower than C.
For Python, the overhead is proportional to the input size and is due to TensorFlow internally having to unbox the Python linked list objects and create a numeric buffer that can be used by the C code.
The unboxing is not needed for NumPy since TensorFlow can use NumPy's internal numeric buffer directly.
By using the C API directly, \carml can elide measuring overheads due to language binding or scripting language use.

\carml{} design supports agents on ASIC and FPGA.
Any code implementing the predictor interface shown in Listing~\ref{lst:predictor_api} is a valid \carml{} predictor.
This means that FPGA and ASIC hardware, which do not have a framework per se, can be exposed as a predictor.
For example, for an FPGA the \texttt{Open} function call loads a bitfile into the FPGA, the \texttt{Close} unloads it, and the \texttt{Predict} runs the inference on the FPGA.
Except for implementing these $3$ API functions, no code needs to change for the FPGA to be exposed to \carml. %

\begin{figure}
  \centering
  \includegraphics[clip, width=0.45\textwidth]{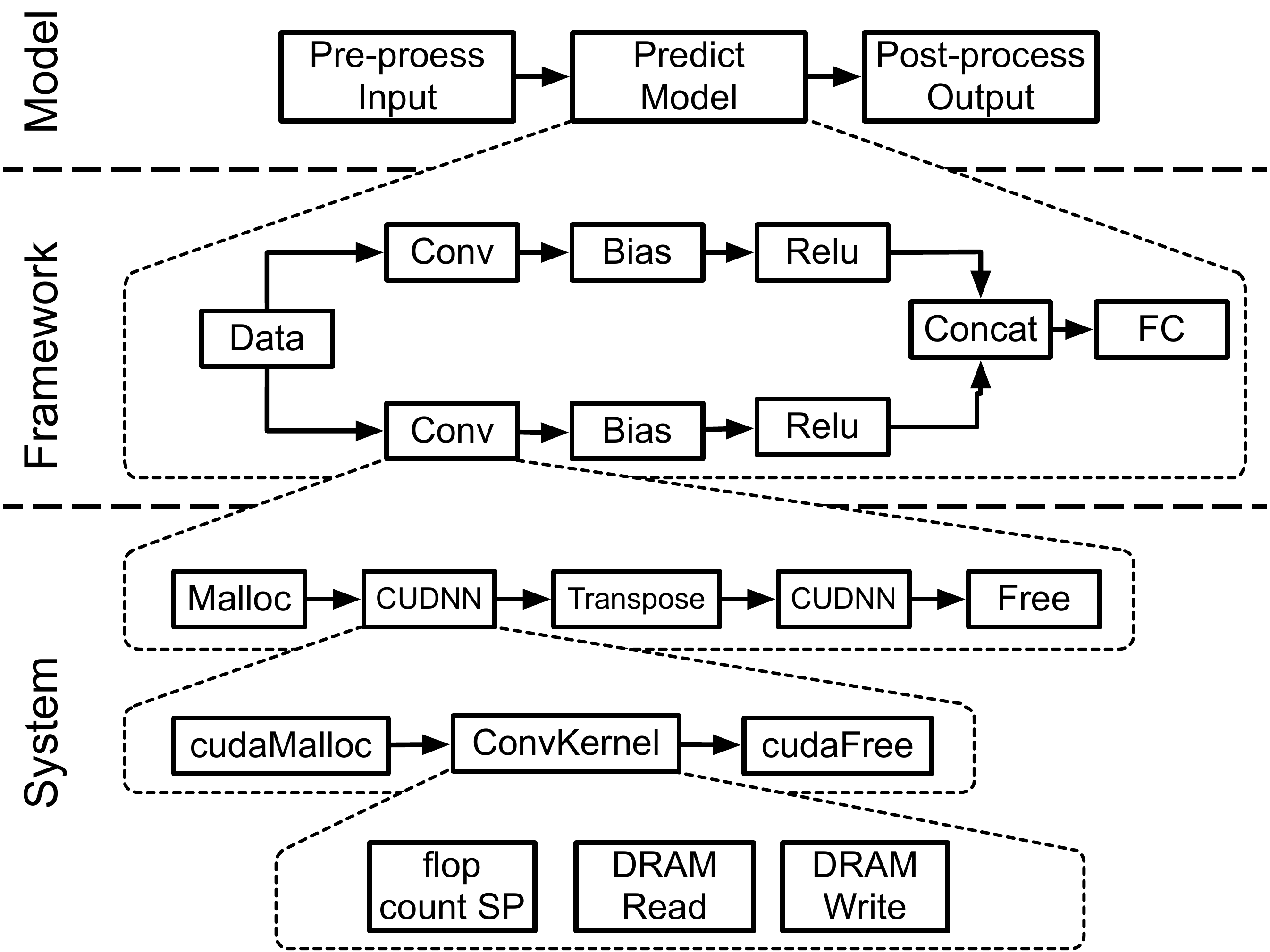}
  \caption{
  The model inference is defined by the pre-processing, prediction, and post-processing pipeline.
  A framework executes a model through a network-layer execution pipeline. Layers executed by a framework are pipelines of system library calls.
  The system libraries, in turn, invoke a chain of primitive kernels that impact the underlying hardware counters.
  }
  \label{fig:profile_levels}
\end{figure}

\subsubsection{Tracing Hooks}\label{sec:tracinghooks}

To enable ~\featureInspection{}, \carml leverages distributed tracing~\cite{sigelman2010dapper} and captures the profiles at different levels of granularity (model-, framework-, and system-level as shown in Figure~\ref{fig:profile_levels}) using the tracing hooks.
A \textit{tracing hook} is a pair of start and end code snippets and follows the standards~\cite{opentracing} to capture an interval of time.
The captured time interval along with the context and metadata is called a \textit{trace event}.
and is published to the \carml{} tracer server (Section~\ref{sec:tracer}).
Trace events are published asynchronously to the \carml tracing server, where they are aggregated using the timestamp and context information into a single end-to-end timeline.
The timestamps of trace events do no need to reflect the actual wall clock time, for example, users may integrate a system simulator and publish simulated time rather than wall-clock time to the tracing server.

\textbf{Model-level} ---
Tracing hooks are automatically placed around each pipeline operator within the model evaluation pipeline.
For example, the tracing hook around the model inference step measures the inference latency.

\textbf{Framework-level} ---
The tracing hooks at the framework-level leverage the DL frameworks' existing profiling capabilities and does not require modification to the framework source code.
For TensorFlow, this option is controlled by the \texttt{RunOptions.TraceLevel} setting which is passed to the \texttt{TF\_SessionRun} function.
In MXNet, the \texttt{MXSetProfilerState} function toggles the layer profiling.
Similar mechanisms exist for other frameworks such as Caffe, Caffe2, PyTorch, and TensorRT.
The framework's profile representation is converted and is then published to the tracer server.

\textbf{System-level} ---
The tracing hooks at the system-level integrate with hardware and system-level profiling libraries to capture fine-grained performance information --- CPU and GPU profiles, system traces, and hardware performance counters.
For example, the performance counters on systems are captured through integration with PAPI~\cite{papi} and Linux perf~\cite{perf} while the GPU profile is captured by integrating with NVIDIA's NVML~\cite{nvml} and CUPTI~\cite{cupti}.
Since overhead can be high for system-level profiling, the user can selectively enable/disable the integrated profilers.

The trace level is a user-specified option (part of the benchmarking scenario) and allows one to get a hierarchical view of the execution profile.
For example, a user can enable model- and framework-level profiling by setting the trace level to \texttt{framework}, or can disable the profiling all together by setting the trace level to none.
Through \carml's trace, a user can get a holistic view of the model evaluation to identify bottlenecks at each level of inference.

%% file: latex_figures/fig_language_cpu.tex
\begin{subfigure}[b]{0.48\textwidth}
\centering
    \setlength{\belowcaptionskip}{0pt}
    \setlength{\abovecaptionskip}{-15pt}
\begin{tikzpicture}
\begin{axis}[
    ymajorgrids=true,
    grid style=dashed,
	ybar interval=0.7,
	xlabel=(b) Normalized CPU Latency.,
	font=\sffamily,
	enlargelimits=0.05,
	tick label style={font=\footnotesize},
	label style={font=\small},
    legend columns=5,
	width=0.8\linewidth,
    height=0.1\linewidth,
    axis x line*=bottom,
    axis y line=left,
    legend style={
    at={(0,0)},anchor=north west,at={(axis description cs:0.26,1.5)},
    font=\sffamily\footnotesize,fill=none,draw=none},
	cycle list name=Dark2-8,
	every axis plot/.append style={fill,draw=none,no markers},
	symbolic x coords = {1,2,4,8,16,32,999},
    xtick=data,
    bar shift=0pt,
    bar width = 10pt,
    ymin=0,
    legend image code/.code={%
        \draw[#1, draw=none] (0cm,-0.1cm) rectangle (0.2cm,0.1cm);
    },
]
	\addplot[plotcolor1] table [x index=0,y index=4] {language_cpu_inceptionv3_tensorflow.dat};
	\addlegendentry{C}

	\addplot[plotcolor3] table [x index=0,y index=5] {language_cpu_inceptionv3_tensorflow.dat};
	\addlegendentry{NumPy}

	\addplot[plotcolor4] table [x index=0,y index=6] {language_cpu_inceptionv3_tensorflow.dat};
	\addlegendentry{Python}

\end{axis}
\end{tikzpicture}
\vspace{-10px}
\label{fig:eval_cpu_language}
\end{subfigure}

%% file: latex_figures/fig_language_gpu.tex
\begin{subfigure}[b]{0.48\textwidth}
\centering
    \setlength{\belowcaptionskip}{0pt}
    \setlength{\abovecaptionskip}{-15pt}
\begin{tikzpicture}
\begin{axis}[
    ymajorgrids=true,
    grid style=dashed,
	ybar interval=0.7,
	xlabel=(b) Normalized GPU Latency.,
	font=\sffamily,
	enlargelimits=0.05,
	tick label style={font=\footnotesize},
	label style={font=\small},
	width=0.8\linewidth,
    height=0.1\linewidth,
    axis x line*=bottom,
    axis y line=left,
	cycle list name=Dark2-8,
	every axis plot/.append style={fill,draw=none,no markers},
	symbolic x coords = {1,2,4,8,16,32,999},
    xtick=data,
    bar shift=0pt,
    bar width = 10pt,
    ymin=0,
    legend columns=5,
    legend style={
    at={(0,0)},anchor=north west,at={(axis description cs:0,1.25)},
    font=\sffamily\footnotesize,fill=none,draw=none},
    legend image code/.code={%
        \draw[#1, draw=none] (0cm,-0.1cm) rectangle (0.2cm,0.1cm);
    },
]
	\addplot[plotcolor1] table [x index=0,y index=4] {language_gpu_inceptionv3_tensorflow.dat};

	\addplot[plotcolor3] table [x index=0,y index=5] {language_gpu_inceptionv3_tensorflow.dat};

	\addplot[plotcolor4] table [x index=0,y index=6] {language_gpu_inceptionv3_tensorflow.dat};

\end{axis}
\end{tikzpicture}
\label{fig:eval_gpu_language}
\end{subfigure}

%% file: sec/4.3-middleware.tex
\subsection{Middleware}\label{sec:middleware}

The \carml middleware layer is composed of services and utilities that support the \carml Server in orchestrating model evaluations and the \carml agents in provisioning, monitoring, and aggregating the execution of the agents.

\subsubsection{Distributed Registry}\label{sec:registry}

\carml leverages a distributed key-value store to store the registered model manifests and running agents, referred to as the \textit{distributed registry}.
\carml uses the registry to facilitate the discovery of models, solve user-specified constraints for selecting \carml{} agents, and load balances the requests across agents.
The registry is dynamic --- both model manifests and predictors can be added or deleted at runtime throughout the lifetime of \carml{}.

\subsubsection{Evaluation Database}\label{sec:database}

In the benchmarking workflow, after completing a model evaluation, the \carml agent uses the user input as the key to store the benchmarking result and profiling trace in the \textit{evaluation database}.
\carml summarizes and generates plots to aid in comparing the performance across experiments.
Users can view historical evaluations through the website or command line using the input constraints.
Since the models are versioned, \carml allows one to track which model version produced the best result.

\subsubsection{Tracing Server}\label{sec:tracer}

The \carml \textit{tracing server} is a distributed tracing server which accepts profiling data published by the \carml{} agent's trace hooks (Section~\ref{sec:tracinghooks}).
Through the innovative use of distributed tracing (originally designed to monitor distributed applications), \carml joins profiling results from different profiling tools and accepts instrumentation markers within application and library code. 
All profiling data are incorporated into a single profiling timeline.
The aggregated profiling trace is consumed by the \carml{} analysis pipeline and also visualized separately where the user can view the entire timeline and  ``zoom'' into a specific component as shown in Figure~\ref{fig:profile_levels}.
As stated in Section~\ref{sec:tracinghooks}, user-specified options control the granularity (model, framework, or system) of the trace events captured (Lines $4-9$ in Listing~\ref{lst:predict_rpc}).

%% file: sec/4.x-extensibility.tex
\subsection{Extensibility and Customization}

\carml is built from a set of modular components and is designed to be extensible and customizable.
Users can disable components, such as tracing, with a runtime option or conditional compilation, for example.
Users can extend \carml by adding models, frameworks, or tracing hooks.

\textbf{Adding Models} --- 
As models are defined through the model manifest file, no coding is required to add models.
Once a model is added to \carml, then it can be used through its website, command line, or API interfaces.
Permissions can be set to control who can use or view a model.

\textbf{Adding Frameworks} ---
To use new or custom versions of a built-in framework requires no code modification but a framework manifest as shown in Listing \ref{lst:framework_manifest}.
To add support for a new type of framework in \carml, the user needs to implement the framework wrapper and expose the framework as a \carml predictor.
The predictor interface is defined by a set of $3$ functions --- one to open a model, another to perform the inference, and finally, one to close the model --- as shown in Listing~\ref{lst:predictor_api}.
The auxiliary code that forms an agent is common across frameworks and does not need to be modified.

\textbf{Adding Tracing Hooks} --- 
\carml is configured to capture a set of default system metrics using the system-level tracing hooks.
Users can configure these existing tracing hooks to capture other system metrics.
For example, to limit profiling overhead, by default, the CUPTI tracing hooks captures only some CUDA runtime API, GPU activities (kernels and memory copy), and GPU metrics.
They can be configured to capture other GPU activities and metrics, or NVTX markers.
Moreover, users can integrate other system profilers into \carml{} by implementing the tracing hook interface (Section~\ref{sec:tracinghooks}).

%% file: sec/5-implementation.tex
\subsection{Implementation}\label{sec:implementation}

We implemented the \carml design
with support for common frameworks and hardware.
At the time of writing, \carml has built-in support for
Caffe, Caffe2, CNTK, MXNet, PyTorch, TensorFlow, TensorFlow Lite, and TensorRT.
\carml{} works with binary versions of the frameworks (version distributed through Python's pip, for example) and support customized versions of the frameworks with no code modification.
\carml has been tested on X86, PowerPC, and ARM CPUs as well as NVIDIA's Kepler, Maxwell, Pascal, Volta, and Turing GPUs.
It can also evaluate models deployed on FPGAs.
During the evaluation, users can specify hardware constraints such as: whether to run on CPU/GPU/FPGA, type of architecture, type of interconnect, and minimum memory requirements --- which \carml uses for agent resolution.

We populated \carml with over $300$ built-in models covering a wide array of inference tasks such as image classification, object detection, segmentation, image enhancement, recommendation, etc. %
We verified \carml's accuracy and performance results by evaluating the built-in models and frameworks across representative systems and comparing to those publicly reported.
We maintain a running version of \carml (omitting the web link due to the blind review process) on a representative set of systems along with the evaluation results of the built-in artifacts.
It serves as a portal for the public to evaluate and measure the systems, and to contribute to \carml's artifacts.
Using the analysis pipeline, we automatically generated profiling reports for hundreds of models across frameworks.
The analysis reports are published as web pages and a sample is available at \resultswebsiteurl for the reader's inspection.

We implemented a \carml web UI using the React Javascript framework.
The web UI interacts with a REST API provided by the server.
A video demoing the web UI usage flow is available at \url{https://bit.ly/2N9Z5wR}.
The REST API can be used by other clients that wish to integrate \carml within their workflow.
A \carml command-line client is also available and can be used within shell scripts.
The agents also expose a gRPC API which can be used to perform queries to the agents directly.

%% file: sec/6-evaluation.tex
\input{sec/6-system_list.tex}

\input{sec/6-model_list.tex}

\section{Evaluation}\label{sec:evaluation}

Previous sections discussed in detail how \carml's design and implementation achieves the \feature{1-6} and \feature{10} design objectives.
In this section, we focus on evaluating how \carml handles  \featureBenchScenarios{} different benchmarking scenarios, \featureReport{} result summarization, and \featureInspection{} inspection of model execution.
We installed \carml{} on the systems listed in Table~\ref{tbl:systems}.
Unless otherwise noted,  all \carml agents are run within a docker container built on top of  NVIDIA's TensorFlow NGC v$19.06$ docker image with the TensorFlow v$1.13.1$ library.
All evaluations were performed using the command-line interface and are run in parallel across the systems.

\input{sec/6.1-modeleval.tex}

\input{sec/6.3-inspection.tex}

\input{sec/6.2-analysis.tex}

%% file: sec/6-system_list.tex
\begin{table*}
    \centering
    \resizebox{0.90\textwidth}{!}{%
        \begin{tabular}{llllrrrr} \toprule
        \thead{\textbf{Name}} & \thead{\textbf{CPU}} & \thead{\textbf{GPU }} & \thead{\shortstack{\textbf{GPU} \\ \textbf{Architecture}}} & \thead{\shortstack{\textbf{GPU Theoretical} \\ \textbf{Flops (TFlops)}}} & \thead{\shortstack{\textbf{GPU Memory} \\ \textbf{Bandwidth (GB/s)}}} & \thead{\shortstack{\textbf{Cost} \\ \textbf{(\$/hr)}}} \\   \midrule
        AWS P3 (2XLarge) & Intel Xeon E5-2686 v4 @ 2.30GHz & Tesla V100-SXM2-16GB & Volta & 15.7 & 900 & 3.06 \\
        AWS G3 (XLarge) & Intel Xeon E5-2686 v4 @ 2.30GHz & Tesla M60  & Maxwell & 9.6 & 320 & 0.90 \\
        AWS P2 (XLarge) & Intel Xeon E5-2686 v4 @ 2.30GHz & Tesla K80 & Kepler & 5.6 & 480 & 0.75 \\  \hdashline
        IBM\ P8 & IBM S822LC Power8 @ 3.5GHz & Tesla P100-SXM2 & Pascal & 10.6 & 732 & - \\
        \bottomrule
        \end{tabular}%
	}%
    \caption{
    Four systems with Volta, Pascal, Maxwell, and Kepler GPUs are selected for evaluation.
    }
    \label{tbl:systems}
\end{table*}

%% file: sec/6-model_list.tex
\begin{table*}
    \centering
    \resizebox{0.90\textwidth}{!}{%
\begin{tabular}{rlcrrrrrrr} \toprule
\centering%
\textbf{ID} & \textbf{Name} & \shortstack{\textbf{Top 1} \\ \textbf{Accuracy}}  &  \shortstack{\textbf{Graph Size} \\ \textbf{(MB)}} & \shortstack{ \textbf{Online} \\  \textbf{TrimmedMean} \\ \textbf{Latency (ms)}} & \shortstack{ \textbf{Online} \\ \textbf{$90^\text{th}$ Percentile} \\ \textbf{Latency (ms)}} & \shortstack{ \textbf{Max Throughput} \\ \textbf{ (Inputs/Sec)}} & \shortstack{ \textbf{Optimal} \\ \textbf{Batch Size}}  \\ \midrule
1 & Inception\_ResNet\_v2 & 80.40 & 214 & 23.95 & 24.2 & 346.6 & 128 \\
2 & Inception\_v4 & 80.20 & 163 & 17.36 & 17.6 & 436.7 & 128\\
3 & Inception\_v3 & 78.00 & 91 & 9.2 & 9.48 & 811.0 & 64  \\
4 & ResNet\_v2\_152 & 77.80 & 231 & 14.44 & 14.65 & 466.8 & 256 \\
5 & ResNet\_v2\_101 & 77.00 & 170 & 10.31 & 10.55 & 671.7 & 256 \\
6 & ResNet\_v1\_152 & 76.80 & 230 & 13.67 & 13.9 & 541.3 & 256 \\
7 & MLPerf\_ResNet50\_v1.5 & 76.46 & 103 & 6.33 & 6.53 & 930.7 & 256 \\
8 & ResNet\_v1\_101 & 76.40 & 170 & 9.93 & 10.08 & 774.7 & 256 \\
9 & AI\_Matrix\_ResNet152 & 75.93 & 230 & 14.58 & 14.72 & 468.0 & 256 \\
10 & ResNet\_v2\_50 & 75.60 & 98 & 6.17 & 6.35 & 1,119.7 & 256 \\
11 & ResNet\_v1\_50 & 75.20 & 98 & 6.31 & 6.41 & 1,284.6 & 256 \\
12 & AI\_Matrix\_ResNet50 & 74.38 & 98 & 6.11 & 6.25 & 1,060.3 & 256 \\
13 & Inception\_v2 & 73.90 & 43 & 6.28 & 6.56 & 2,032.0 & 128 \\
14 & AI\_Matrix\_DenseNet121 & 73.29 & 31 & 11.17 & 11.49 & 846.4 & 32 \\
15 & MLPerf\_MobileNet\_v1 & 71.68 & 17 & 2.46 & 2.66 & 2,576.4 & 128 \\
16 & VGG16 & 71.50 & 528 & 22.43 & 22.59 & 687.5 & 256 \\
17 & VGG19 & 71.10 & 548 & 23.0 & 23.31 & 593.4 & 256 \\
18 & MobileNet\_v1\_1.0\_224 & 70.90 & 16 & 2.59 & 2.75 & 2,580.6 & 128 \\
19 & AI\_Matrix\_GoogleNet & 70.01 & 27 & 5.43 & 5.55 & 2,464.5 & 128 \\
20 & MobileNet\_v1\_1.0\_192 & 70.00 & 16 & 2.55 & 2.67 & 3,460.8 & 128 \\
21 & Inception\_v1 & 69.80 & 26 & 5.27 & 5.41 & 2,576.6 & 128 \\
22 & BVLC\_GoogLeNet & 68.70 & 27 & 6.05 & 6.17 & 951.7 & 8 \\
23 & MobileNet\_v1\_0.75\_224 & 68.40 & 10 & 2.48 & 2.61 & 3,183.7 & 64 \\
24 & MobileNet\_v1\_1.0\_160 & 68.00 & 16 & 2.57 & 2.74 & 4,240.5 & 64 \\
25 & MobileNet\_v1\_0.75\_192 & 67.20 & 10 & 2.42 & 2.6 & 4,187.8 & 64 \\
26 & MobileNet\_v1\_0.75\_160 & 65.30 & 10 & 2.48 & 2.65 & 5,569.6 & 64 \\
27 & MobileNet\_v1\_1.0\_128 & 65.20 & 16 & 2.29 & 2.46 & 6,743.2 & 64 \\
28 & MobileNet\_v1\_0.5\_224 & 63.30 & 5.2 & 2.39 & 2.58 & 3,346.5 & 64 \\
29 & MobileNet\_v1\_0.75\_128 & 62.10 & 10 & 2.3 & 2.47 & 8,378.4 & 64 \\
30 & MobileNet\_v1\_0.5\_192 & 61.70 & 5.2 & 2.48 & 2.67 & 4,453.2 & 64 \\
31 & MobileNet\_v1\_0.5\_160 & 59.10 & 5.2 & 2.42 & 2.58 & 6,148.7 & 64 \\
32 & BVLC\_AlexNet & 57.10 & 233 & 2.33 & 2.5 & 2,495.8 & 64 \\
33 & MobileNet\_v1\_0.5\_128 & 56.30 & 5.2 & 2.21 & 2.33 & 8,924.0 & 64 \\
34 & MobileNet\_v1\_0.25\_224 & 49.80 & 1.9 & 2.46 & 3.40 & 5,257.9 & 64 \\
35 & MobileNet\_v1\_0.25\_192 & 47.70 & 1.9 & 2.44 & 2.6 & 7,135.7 & 64 \\
36 & MobileNet\_v1\_0.25\_160 & 45.50 & 1.9 & 2.39 & 2.53 & 10,081.5 & 256 \\
37 & MobileNet\_v1\_0.25\_128 & 41.50 & 1.9 & 2.28 & 2.46 & 10,707.6 & 256 \\
\bottomrule
\end{tabular}%
	}
    \caption{$37$  pre-trained TensorFlow image classification models from MLPerf~\cite{mlperf}, AI-Matrix~\cite{aimatrix}, and TensorFlow Slim are used for evaluation and are sorted by  accuracy.
    The graph size is the size of the frozen graph for a model.
    We measured the online latency, $90^\text{th}$ percentile latency, maximum throughput in batched inference at the optimal batch size for each model.
    }
    \label{tab:tf_models}
\end{table*}

%% file: sec/6.1-modeleval.tex
\subsection{Benchmarking Scenarios}\label{sec:eval:models}

To show how \carml helps users choose from different models and system offerings for the same DL task, we compared the inference performance across the $37$ TensorFlow models (shown in Table~\ref{tab:tf_models}) and systems (shown in Table~\ref{tbl:systems}) under different benchmark scenarios.
For each model, we measured its trimmed mean latency~\footnote{Trimmed mean is computed by removing $20\%$ of the smallest and largest elements and computing the mean of the residual; i.e. $\textbf{\texttt TrimmedMean} \left ( list) \right ) = \textbf{\texttt Mean} ( \textbf{\texttt Sort}\left(list\right )$ $[\lfloor 0.2*\textbf{\texttt len}(list)\rfloor \text{\texttt ::}-\lfloor 0.2*\textbf{\texttt len}(list)\rfloor ] )$.} and $90^\text{th}$ percentile latency in online (batch size = $1$) inference scenario, and the maximum throughput in batched inference scenario on the \texttt{AWS P3} system in Table~\ref{tbl:systems}.
The model accuracy achieved using the ImageNet~\cite{deng2009imagenet} validation dataset and model size is listed.
A model deployer can use this accuracy and performance information to choose the best model on a system given the accuracy and target latency or throughput objectives.

\begin{figure*}
  \centering
   \begin{minipage}{0.48\textwidth}
   \includegraphics[width=\textwidth]{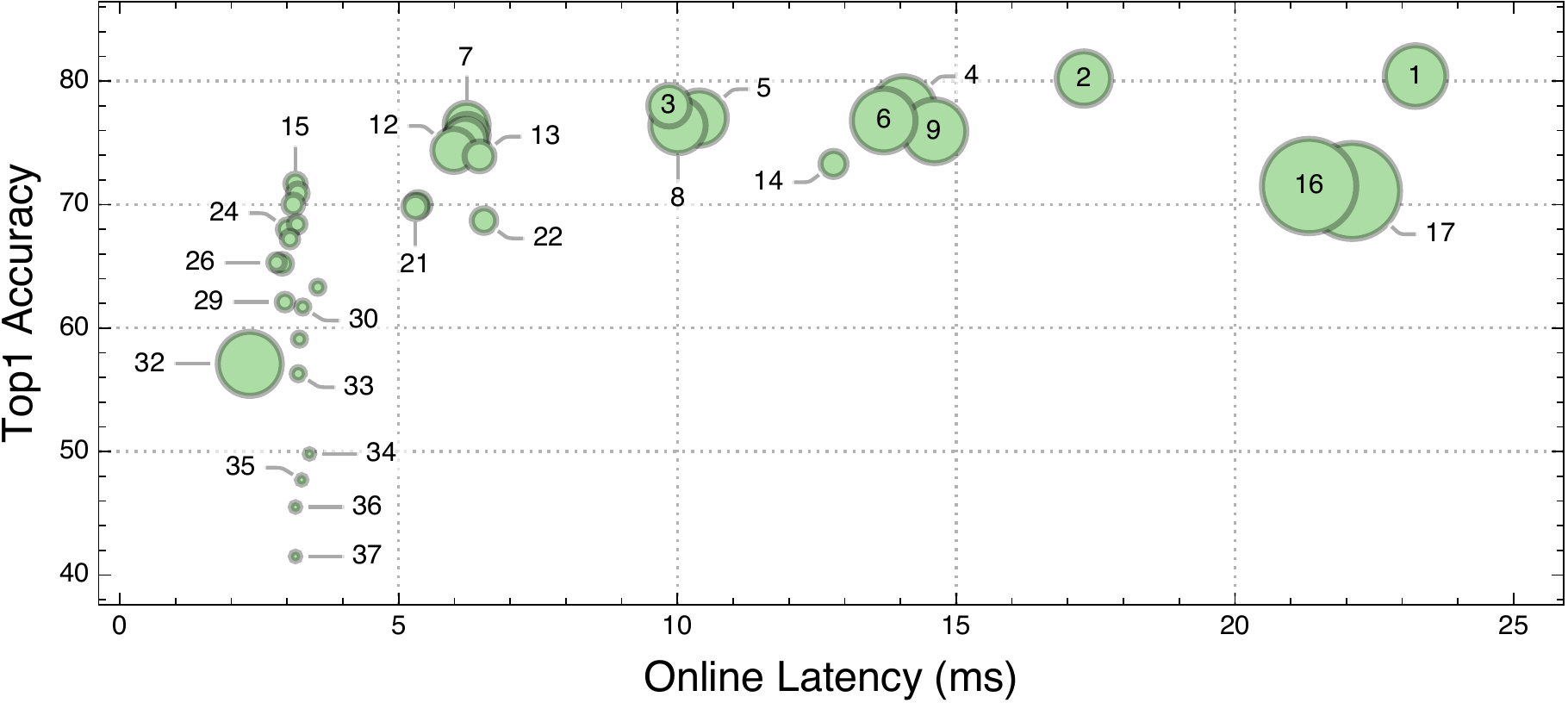}
   \caption{Accuracy vs online latency on AWS P3.}
        \label{fig:accuracy_latency_compare}
    \end{minipage}
    \hfill
   \begin{minipage}{0.48\textwidth}
    \includegraphics[width=\textwidth]{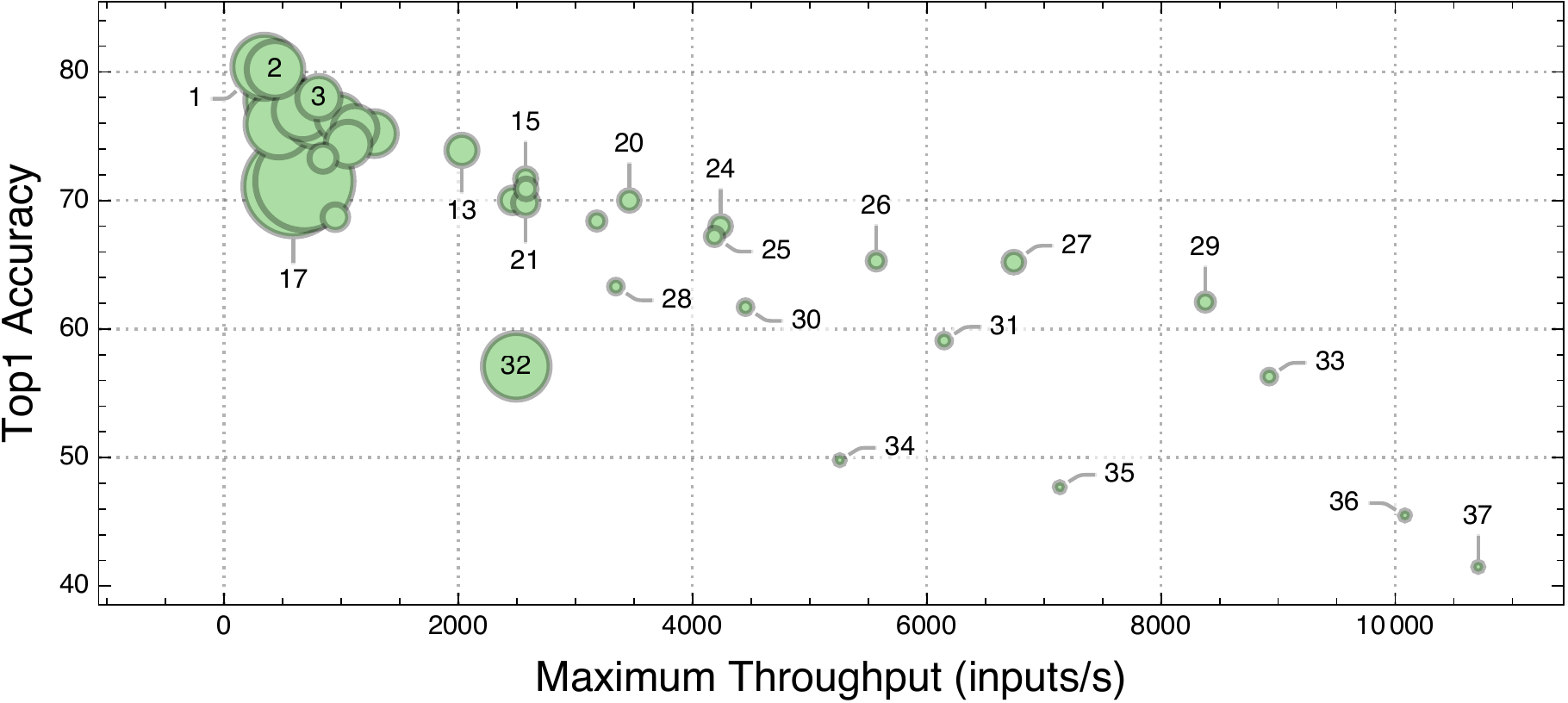}
    \caption{Accuracy vs maximum throughput on AWS P3.}
    \label{fig:accuracy_throughput_compare}
    \end{minipage}
\end{figure*}

\textbf{Model Accuracy, Size, and Performance} --- 
We examined the relationship between the model accuracy and both online latency (Figure~\ref{fig:accuracy_throughput_compare}) and maximum throughput (Figure \ref{fig:accuracy_latency_compare}).
In both figures, the area of the circles is proportional to the model's graph size.
In Figure \ref{fig:accuracy_latency_compare} we find a limited correlation between a model's online latency and its accuracy --- models taking longer time to run do not necessarily achieve higher accuracies; e.g. model $15$ vs $22$.
While large models tend to have longer online latencies, this is not always true; e.g. model $14$ is smaller in size but takes longer to run compared to models $3$, $5$, $8$, etc.
Similarly, in Figure \ref{fig:accuracy_throughput_compare}, we find a limited correlation between a model's accuracy and its maximum throughput --- two models with comparable maximum throughputs can achieve quite different accuracies; e.g. models $2$ and $17$.
Moreover, we see from both figures that the graph size (which roughly represents the number of weight values) is not directly correlated to either accuracy or performance.
Overall, models closer to the upper left corner (low latency and high accuracy) in Figure~\ref{fig:accuracy_latency_compare} are favorable in the online inference scenarios, and models closer to the upper right corner (high throughput and high accuracy) in  Figure~\ref{fig:accuracy_throughput_compare} are favorable in the batched inference scenario.
Users can use this information to select the best model depending on their objectives.

\begin{figure}
  \centering
  \includegraphics[width=0.48\textwidth]{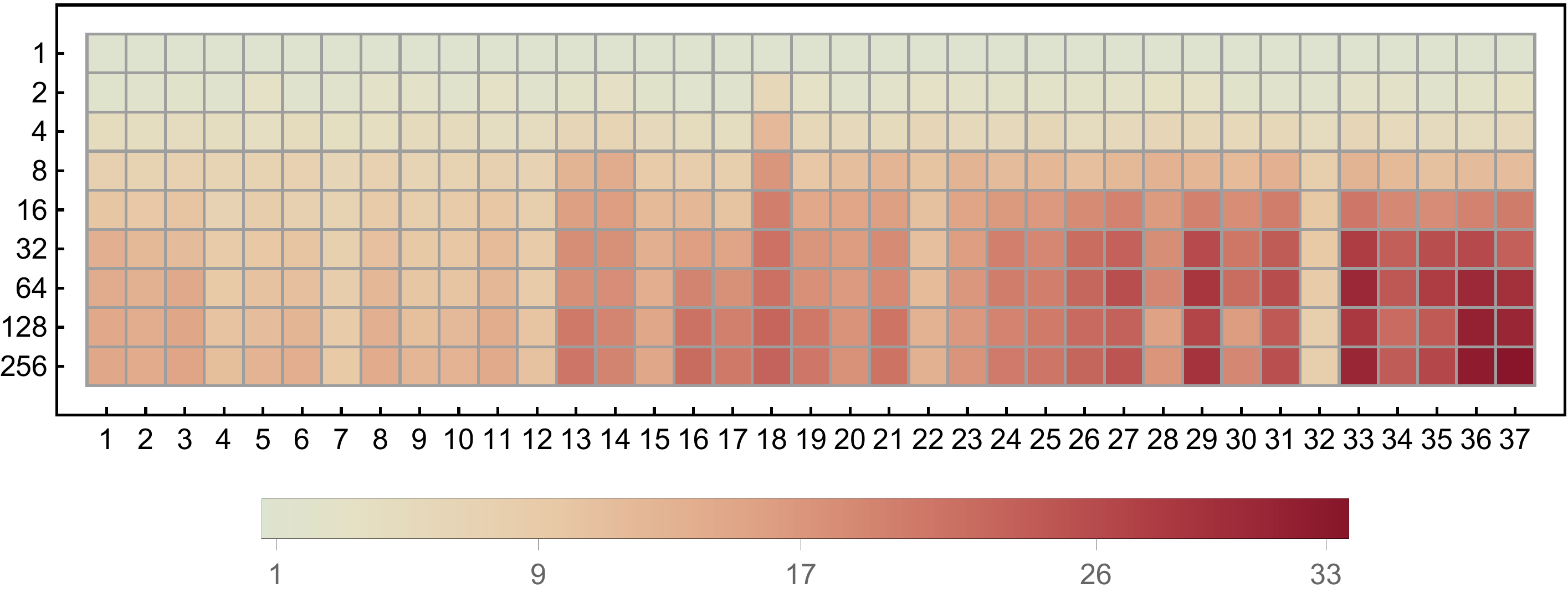}
  \caption{The throughput speedup (over batch size 1) heatmap across batch sizes on AWS P3 for the $37$ models in Table~\ref{tab:tf_models}. The $y-$axis shows the batch size, whereas the $x-$axis shows the model ID. 
  }
  \label{fig:batchsize_speedup}
\end{figure}

\begin{figure}
  \centering
  \includegraphics[width=0.48\textwidth]{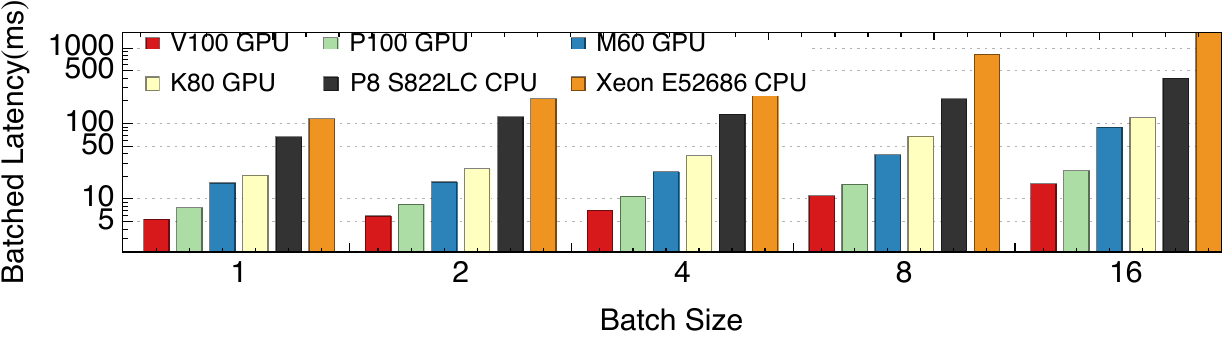}
  \caption{The batched latency of \texttt{ResetNet 50} across the GPUs and CPUs listed in Table~\ref{tbl:systems}.
  }
  \label{fig:resnet50_latencies}
\end{figure}

\textbf{Model Throughput Scalability Across Batch Sizes} --- 
When comparing the model online latency and maximum throughput (Figures~\ref{fig:accuracy_latency_compare} and \ref{fig:accuracy_throughput_compare} respectively), we observed that models which exhibit good online inference latency do not necessarily perform well in the batched inference scenario where throughput is important.
We measured how model throughput scales with batch size (referred to as \textit{throughput scalability}) and present this model characteristic in Figure~\ref{fig:batchsize_speedup}.
As shown, the throughput scalability varies across models.
Even models with similar network architectures can have different throughput scalability --- e.g. models $4$ and $6$, models $5$ and $8$, and models $10$ and $11$.
In general, smaller models tend to have better throughput scalability.
However, there are exceptions, for example, the VGG models ($16$ and $17$) are large and have good throughput scalability. %

\textbf{Model Performance Across Systems} --- 
Overall, the \texttt{ResNet\_50} class of models offer a balance between model size, accuracy, performance and are commonly used in practice. 
Thus we use \texttt{ResNet\_50} in online inference as an example to show how to use \carml{} to choose the best system given a model.
We evaluated \texttt{ResNet\_50} across all CPUs and GPUs listed in Table~\ref{tbl:systems} and the results are shown in Figure~\ref{fig:resnet50_latencies}.
On the CPU side, IBM S822LC Power8 achieves between $1.7\times$ and $4.1\times$ speedup over Intel Xeon E5-2686.
The P8 CPU is more performant than Xeon CPU~\cite{elisseev2018study}, with the P8 running at $3.5$ GHz and having $10$ cores each capable of running $80$ SMT threads.
On the GPU side, as expected, V100 GPU achieves the lowest latency followed by the P100.
The M60 GPU is $1.2\times$ to $1.7\times$ faster than the K80.
When this information is coupled with the pricing information of the systems, one can 
determine which system is most cost-efficient given a latency target and benchmarking scenario.
For example, given that K80 costs $0.90$\$/hr  and M60 costs $0.75$\$/hr on AWS, we can tell that M60 is both more cost-efficient and faster than K80 --- thus, M60 is overall better suited for \texttt{ResNet\_50} online inference when compared to K80 on AWS.

%% file: sec/6.3-inspection.tex
\begin{table*}
    \centering
    \resizebox{0.90\textwidth}{!}{%
        \begin{tabular}{rllrlrr} \toprule
        \centering%
        \thead{\shortstack{\textbf{Layer} \\ \textbf{Index}}} & \thead{\shortstack{\textbf{Layer} \\ \textbf{Name}}} & \thead{\shortstack{\textbf{Layer} \\ \textbf{Type}}} & \thead{\shortstack{\textbf{Layer Shape}}}  & \thead{\shortstack{\textbf{Dominant GPU Kernel(s) Name}}}  & \thead{\shortstack{\textbf{Latency} \\ \textbf{(ms)}}} & \thead{\shortstack{\textbf{Alloc Mem} \\ \textbf{(MB)}}} \\   \midrule
        208 & \texttt{conv2d\_48/Conv2D}	& Conv2D & $\langle 256, 512, 7, 7 \rangle$ & \texttt{volta\_cgemm\_32x32\_tn} & 7.59 & 25.7 \\
        221 & \texttt{conv2d\_51/Conv2D} & Conv2D & $\langle 256, 512, 7, 7 \rangle$ & \texttt{volta\_cgemm\_32x32\_tn} &  7.57 & 25.7 \\
        195 & \texttt{conv2d\_45/Conv2D}	& Conv2D & $\langle 256, 512, 7, 7 \rangle$ & \texttt{volta\_scudnn\_128x128\_relu\_interior\_nn\_v1} & 5.67 & 25.7 \\
        3 & \texttt{conv2d/Conv2D} & Conv2D & $\langle 256, 64, 112, 112 \rangle$ & \texttt{volta\_scudnn\_128x64\_relu\_interior\_nn\_v1} & 5.08 & 822.1 \\
        113 & \texttt{conv2d\_26/Conv2D}	& Conv2D & $\langle 256, 256, 14, 14 \rangle$ & \texttt{volta\_scudnn\_128x64\_relu\_interior\_nn\_v1} & 4.67 & 51.4 \\
        \bottomrule
        \end{tabular}%
	}
    \caption{The ResNet 50 layer information using AWS P3 (Tesla V100 GPU) with batch size 256. The top $5$ most time-consuming layers are summarized from the tracing profile. In total, there are $234$ layers of which $143$ take less than $1$ms.}
    \label{tab:layer_info}
\end{table*}

\subsection{Model Execution Inspection}\label{sec:eval:systems}

\carml's evaluation inspection capability helps users to understand the model execution and identify performance bottlenecks.
We show this by performing a case study of  ``cold-start'' inference (where the model needs to be loaded into the memory before inference) of \texttt{BVLC\_AlexNet} (ID = $32$).
The cold-start inference is common on low-memory systems and in serving schemes that perform one-off evaluation (thus models do not persist in memory).

We choose \texttt{BVLC\_AlexNet} because it is easy to see the effects of the ``cold-start'' inference scenario using Caffe on the AWS P3 and IBM P8 GPU systems with batch size $64$.
The results are shown in Figure~\ref{fig:fc_layer}.
We see that IBM P8 with P100 GPU is more performant than AWS P3 which has V100 GPU.
We used \carml's model execution inspection capability to delve deeper into the model and to reveal the reason.
We ``zoomed'' into the longest-running layer (\texttt{fc6}) and find that most of the time is spent performing copies for the (\texttt{fc6}) layer weights.
On AWS P3, the \texttt{fc6} layer takes $39.44ms$ whereas it takes $32.4ms$ on IBM P8.
This is due to the IBM P8 system having an NVLink interconnect which has a theoretical peak CPU to GPU bandwidth of $40$ GB/s ($33$ GB/s measured) while the AWS P3 system performs the copy over PCIe-3 which has a maximum theoretical bandwidth of $16$ GB/s ($12$ GB/s measured). 
Therefore, despite P3's lower compute latency, we observed a lower overall layer and model latency on the IBM P8 system due to the \texttt{fc6} layer being memory bound.

Using \carml's model execution inspection, it is clear that the memory copy is the bottleneck for the ``cold-start'' inference.
To verify this observation, we examined the Caffe source code.
Caffe performs lazy memory copies for layer weights just before execution. 
This causes compute to stall while the weights are being copied --- since the weights of the FC layer are the biggest. %
A better strategy --- used by Caffe2, MXNet, TensorFlow, and TensorRT ---  is to eagerly copy data asynchronously and utilize CUDA streams to overlap compute with memory copies.

%% file: sec/6.2-analysis.tex
\begin{figure}
  \includegraphics[width=0.48\textwidth]{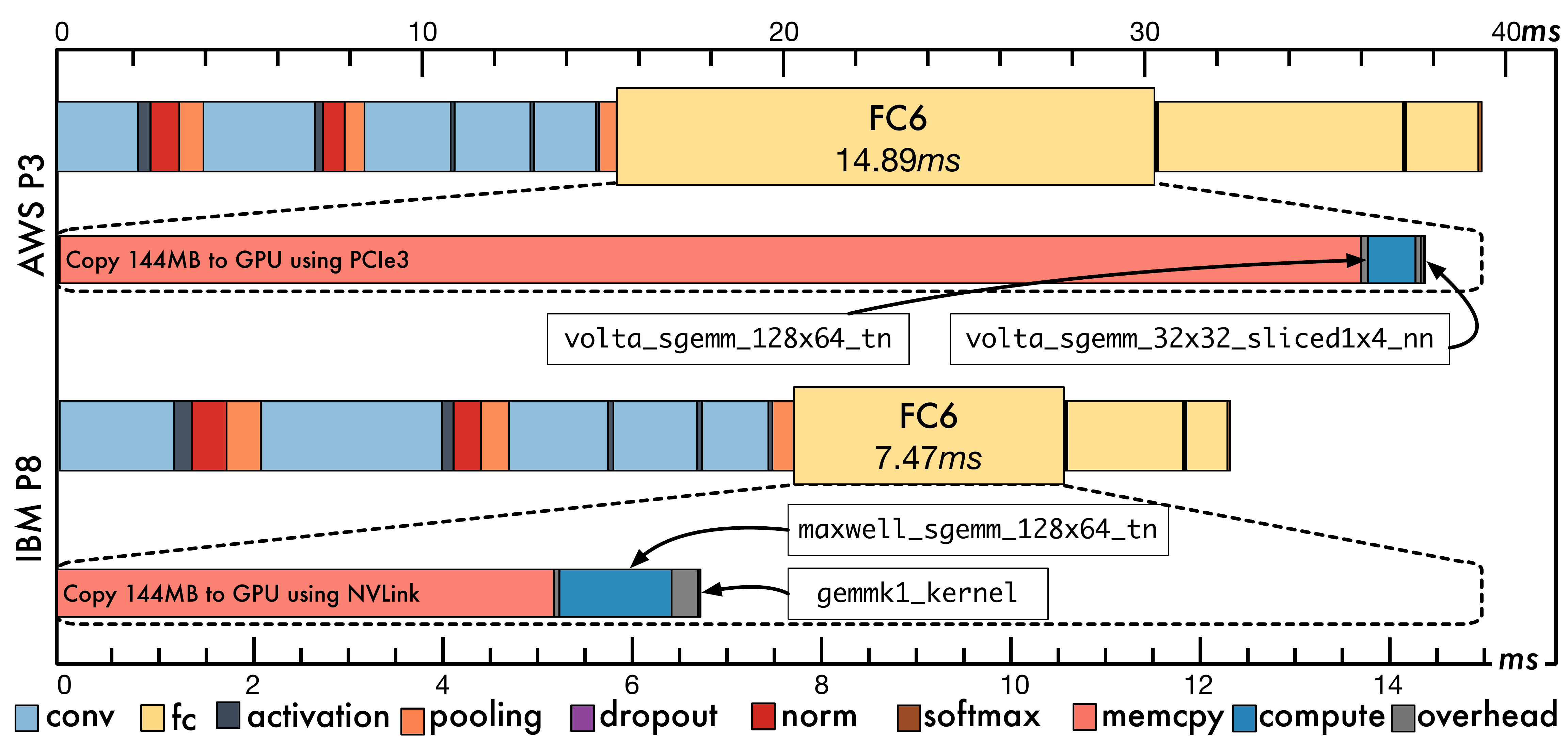}
  \caption{
  The \carml inspection of ``cold-start'' \texttt{BVLC\_AlexNet} inference with batch size $64$ running Caffe v$0.8$ using GPU on AWS P3 and IBM P8 (Table~\ref{tbl:systems}).
  The color-coding of layers signify that they have the same type but does not imply that the layer parameters are the same.
  }
  \label{fig:fc_layer}
\end{figure}

\subsection{Benchmarking Analysis and Reporting}\label{sec:eval:analysis}

To show \carml{}'s benchmarking analysis and reporting capability, we used \carml's analysis workflow to perform an in-depth analysis of the $37$ models.
All results were generated automatically using \carml and further results are available at \resultswebsiteurl for the reader's inspection.
As an example, we highlight the model-layer-GPU kernel analysis of \texttt{ResNet\_50} using batch size $256$ (the optimal batch size with the maximum throughput) on AWS P3.
\carml can capture the layers in a model and correlate the GPU kernels calls to each layer; i.e. tell which GPU kernels are executed by a certain layer.
For example, layer index $208$ is the most time-consuming layer within the model and $7$ GPU kernels are launched by this layer: \kernel{1} \texttt{volta\_cgemm\_32x32\_tn} taking $6.03$ ms,
\kernel{2} \texttt{flip\_filter} taking $0.43$ ms,
\kernel{3} \texttt{fft2d\_r2c\_16x16} taking $0.42$ ms,
\kernel{4} \texttt{fft2d\_c2r\_16x16} taking $0.25$ ms,
\kernel{5} \texttt{fft2d\_r2c\_16x16} taking $0.25$ ms,
\kernel{6} \texttt{ShuffleInTensor3Simple} taking $0.06$ ms, and
\kernel{7} \texttt{compute\_gemm\_pointers} taking $0.004$ ms.
\kernel{1-5} and \kernel{7} are launched by the cuDNN to perform convolution using the FFT algorithm~\cite{jorda2019performance}.
\kernel{6} is launched by TensorFlow and shuffles a layer shape based on a permutation and is used by the TensorFlow convolution layer to convert from TensorFlow's filter format\ignore{ (\texttt{OIHW} -- output channels, input channels, height width)} to the cuDNN filter format\ignore{ (\texttt{NHWC})}.
Table~\ref{tab:layer_info} shows the top $5$ most time-consuming layers of \texttt{ResNet\_50} as well as the dominant kernel (the kernel with the highest latency) within each layer.
Through the analysis and summarization workflow, users can easily digest the results and identify understand model-, framework-, and system-level bottlenecks.

%% file: sec/2.x-related.tex
\section{Related Work}\label{sec:related}

To the authors' knowledge, this the first paper to describe the design and implementation of a scalable DL benchmarking platform.
While there have been efforts to develop certain aspects of \carml, the efforts have been quite dispersed and there has not been a cohesive system that addresses \feature{1-10}.
For example, while there is active work on proposing benchmark suites, reference workloads, and analysis~\cite{mlperf,zhu2018tbd}, they provide \feature{7} a set of benchmarking scenarios and a simple mechanism for \feature{8} analysis and reporting of the results.
The models within these benchmarks can be consumed by \carml, and we have shown analysis which uses the benchmark-provided models.
Other work are purely model serving platforms~\cite{olston2017tensorflow,crankshaw2017clipper} which address \feature{4} scalable evaluation and possibly \feature{5} artifact versioning but nothing else.
Finally, systems such as as~\cite{tsay2018runway,novella2018container,ghanta2018systems} track the model and data from their use in training til deployment and can achieve \feature{1} reproducible and \feature{2} consistent evaluation.

To our knowledge, the most relevant work to \carml is FAI-PEP~\cite{faipep}.
FAI-PEP is a DL benchmarking platform targeted towards mobile devices.
FAI-PEP aims to solve~\feature{1--5} and has limited support of \feature{8} (limited to computing the $n^\text{th}$ percentile latency and displaying plot of these analyzed latencies).
No in-depth profiling and analysis are available within their platform.

%% file: sec/7-conclusion.tex
\section{Conclusion and Future Work}\label{sec:conclusion}

A big hurdle in adopting DL innovations is to evaluate, analyze, and compare their performance.
This paper first identified $10$ design objectives of a DL benchmarking platform.
It then described the design and implementation of \carml --- an open-source DL benchmarking platform that achieves these design objectives.
\carml offers a unified and holistic way to evaluate and inspect DL models, and provides an automated analysis and reporting workflow to summarize the results.
We demonstrated \carml by using it to evaluate a set of models and show how model, hardware, and framework selection affects model accuracy and performance under different benchmarking scenarios.
We are actively working on curating automated analysis and reports obtained through \carml, and a sample of the generated reports is available at \resultswebsiteurl for the reader's inspection.
We are further working on maintaining an online public instance of \carml where users can perform the analysis presented without instantiating \carml on their system.

%% file: sec/99-ack.tex
\begin{acks}

This work is supported by IBM-ILLINOIS Center for Cognitive Computing Systems Research (C3SR) - a research collaboration as part of the IBM Cognitive Horizon Network.

\end{acks}